\documentclass[prc,aps,preprintnumbers,amsmath,amssymb,superscriptaddress,showpacs,floatfix,nofootinbib]{revtex4-1}
\pdfoutput=1

\usepackage{dcolumn}
\usepackage{bm}
\usepackage{amssymb}
\usepackage{amsmath}
\usepackage{ifpdf}
\usepackage{pdfpages}

\newcommand{\be}{\begin{equation}}
\newcommand{\ee}{\end{equation}}
\newcommand{\bea}{\begin{eqnarray}}
\newcommand{\eea}{\end{eqnarray}}

\begin{document}

\title{Sensitivity of flow harmonics to sub-nucleon scale fluctuations in heavy ion collisions}

\author{Jacquelyn Noronha-Hostler}
\affiliation{Department of Physics, Columbia University, 538 West 120th Street, New York,
NY 10027, USA}
\author{Jorge Noronha}
\affiliation{Department of Physics, Columbia University, 538 West 120th Street, New York,
NY 10027, USA}
\affiliation{Instituto de F\'{\i}sica, Universidade de S\~{a}o Paulo, C.P. 66318,
05315-970 S\~{a}o Paulo, SP, Brazil}
\author{Miklos Gyulassy}
\affiliation{Department of Physics, Columbia University, 538 West 120th Street, New York,
NY 10027, USA}

\date{\today}

\begin{abstract}
In this paper a new procedure to smooth out the initial energy densities of hydrodynamics is employed to show that  the initial spatial eccentricities $\varepsilon_{m,n}$, which drive the final state flow harmonics $v_n$, are remarkably robust with respect to variations of the underlying scale of initial energy density spatial gradients, $\lambda$, in nucleus-nucleus collisions. For $\sqrt{s}=2.76$ TeV Pb+Pb collisions, the $\varepsilon_{m,n}$'s (across centrality classes) change by less than $10\%$ if the scale of fluctuations is varied from $0.1$ to $1$ fm. We show, using the 2+1 Lagrangian hydrodynamic code, v-USPhydro, that this robustness is transferred to the final $v_n$'s computed within event by event viscous hydrodynamics. This indicates that the flow harmonics in nucleus-nucleus collisions are not yet particularly sensitive to the underlying microscopic sub-nucleon physics below the confinement scale. On the other hand, the eccentricities of top $1\%$ high multiplicity $\sqrt{s}=5.02$ TeV p+Pb collisions are found to be very sensitive to sub-nucleonic scale fluctuations, which should be contrasted with the robustness found in peripheral Pb+Pb collisions with the same multiplicity.

\end{abstract}   
   

 
\maketitle

\section{Introduction: A tale of scales in nucleus-nucleus collisions}\label{intro}

After more than a decade of intense investigation there is by now large experimental evidence that the local initial energy density of the matter formed in ultrarelativistic heavy ion collisions is sufficiently large enough to produce a deconfined plasma of quarks and gluons \cite{expQGP1,expQGP2,expQGP3,expQGP4,gm}. The main experimental signatures of this new state of matter can be divided into two groups. On one hand, the quark-gluon plasma is color opaque with respect to the propagation of partonic jets, which leads to the experimentally observed suppression of high $p_T$ single inclusive hadrons known as jet quenching \cite{Bjorken:1982tu,Gyulassy:1990ye,Wang:1991xy} (see \cite{Burke:2013yra} for a recent theoretical discussion). On the other hand, 
the large anisotropic flow coefficients of low transverse momentum hadrons, $p_T < 2$ GeV, indicate that there is a significant degree of collectivity in the system, which is generally consistent with relativistic hydrodynamic calculations in the ``perfect fluid" limit where viscous effects are minimal (for a recent review see \cite{heinzsnellings}). This suggests that the non-Abelian quantum fields present in the initial stage of the collisions may behave incoherently at length scales (at least) on the order of the size of a large nucleus in a way that is compatible with the strong assumptions behind relativistic hydrodynamic behavior \cite{Mota:2012qv,deSouza:2015ena}.

However, just as the initial inhomogeneities in the early universe led to the formation of large scale structures after expansion \cite{weinbergcosmo}, the hot and dense matter created in heavy ion collisions displays energy density fluctuations of different characteristic sizes and physical origins. For instance, in the context of heavy ion collisions one may consider the size of a large lead nucleus $R_{Pb}\sim 7-10$ fm as a \emph{macroscopic} nuclear length, the size of a nucleon $\sim 1$ fm as an intermediate \emph{mesoscopic} scale, while smaller length scales of the size of the inverse of the saturation scale $\sim 1/Q_s < 0.1$ fm can be considered as the \emph{microscopic} sub-nucleon regime. The macroscopic nuclear scale is essentially ``geometrical" in the sense that it can be varied by changing the type of nucleus, the final hadron multiplicity yield, and the center-of-mass collision energy per nucleon. The mesoscopic length scale is of the order of the inverse nonperturbative $1/\Lambda_{QCD}$ scale and at that scale one cannot yet resolve the color field configurations inside the nucleons. Finally, the microscopic sub-nucleon scale resolves the internal color structure of the nucleon and some of its features can be understood at weak coupling (though in a nonperturbative manner, for a review see \cite{Gelis:2010nm}). Therefore, ultrarelativistic collisions of heavy nuclei involve QCD phenomena associated with characteristic length scales that can vary by approximately three orders of magnitude. At the moment, there is no effective theory that can describe phenomena at these different scales in a single, consistent framework.

While variations in the macroscopic nuclear scale have been studied at length both theoretically and experimentally over the years \cite{heinzsnellings}, the interplay between mesoscopic and microscopic sub-nucleon scales and their effects in the calculation of heavy ion collision observables, such as the anisotropic flow coefficients, still require further investigation. Given the success of relativistic hydrodynamics to describe anisotropic flow data in nucleus-nucleus collisions, it is reasonable to assume that hydrodynamics may be an adequate effective theory framework to be used in such a study.

Current numerical simulations of relativistic hydrodynamic employ Israel-Stewart-like models \cite{Israel:1979wp} where the shear viscous contribution to the energy-momentum tensor, $\pi^{\mu\nu}$, obeys an independent differential equation of the relaxation type\footnote{This implies that in these models one needs to provide not only initial conditions for the standard hydrodynamic fields, such as the energy density $\epsilon$ (or pressure $P$) and flow velocity $u^\mu$, but also $\pi^{\mu\nu}$ must be independently known at the initial time $\tau_0$. Note that in this paper we do not consider effects from other conserved charges such as baryon number.}. This relaxation process towards the universal Navier-Stokes regime is characterized by the shear relaxation time transport coefficient $\tau_\pi$. For dilute gases described by the relativistic Boltzmann equation, the 14-moments approximation for a massless, single component gas (assuming classical statistics) with constant cross section gives \cite{Denicol:2010xn,Denicol:2011fa}
\be
\tau_\pi = \frac{5}{T}\frac{\eta}{s}\,,
\ee
where $T$ is the temperature, $\eta$ is the shear viscosity, and $s$ is the local entropy density. For instance, if we assume that the largest temperature achieved in central Pb+Pb collisions at the LHC is, say $T \sim 400$ MeV, for $\eta/s=0.1$ the formula above gives $\tau_\pi \sim 0.25$ fm. An important aspect of this formula is that even for a constant $\eta/s$ the relaxation time may vary significantly in space and time according to the local temperature profile.

However, as discussed in \cite{Denicol:2011fa}, in Israel-Stewart (IS) models $\tau_\pi$ is the relevant \emph{microscopic} scale associated with shear stress and, as such, it provides a lower bound on the scales at which one expects this hydrodynamical model to provide accurate results\footnote{Clearly, if the bulk viscosity contribution to the energy-momentum tensor, $\Pi$, is included in the dynamics one has to take into account its corresponding relaxation time $\tau_\Pi$ as well.}. In fact, microscopic sub-nucleon processes that involve scales shorter than $\tau_\pi$ are not properly taken into account in IS models and their description requires the introduction of other degrees of freedom beyond those already present in IS theory. This has been shown to be the case in \cite{Denicol:2012cn}, in a systematic manner, using the relativistic Boltzmann equation as the underlying microscopic theory\footnote{At strong coupling the equations of motion that describe the evolution of the shear stress tensor $\pi^{\mu\nu}$ are not of relaxation-type \cite{Denicol:2011fa,Noronha:2011fi,Heller:2014wfa} and, in this case, IS theory fails to properly describe transient fluid dynamical phenomena.}. Alternatively, the current success of anisotropic hydrodynamics in describing strongly dissipative processes with large momentum space anisotropy requires the introduction of new degrees of freedom in the dynamics that are not the usual hydrodynamic fields (for a recent study, and references therein, see \cite{Nopoush:2015yga}). Therefore, IS-like theories do not correctly resolve scales below $\tau_\pi$ and this intrinsic limitation of this effective theory must be taken into account in numerical simulations. In fact, such a limitation becomes especially relevant in event by event hydrodynamic simulations in which the hydrodynamic fields may possess large spatial gradients at early times.

The presence of large spatial gradients can be verified by computing the Knudsen number, $K_n$, which is roughly speaking the product of the relevant microscopic scale (in our case, $\tau_\pi$) and a spatial gradient of the hydrodynamic fields (in this sense, it is clear that $K_n$ should be small for hydrodynamics to be valid). Using the gradients of $\epsilon$ and $u^\mu$ one can construct a handful of different Knudsen numbers (or fields since these quantities vary in space and time). In \cite{Niemi:2014wta} the regions where $K_n = 0.5$ were considered to define the edge of the validity of the hydrodynamic description and it was found that the Knudsen number $K_{n\theta} = \tau_\pi\, \theta$ associated with the local expansion rate, $\theta  = \nabla_\mu u^\mu$, and the Knudsen number $K_{n\epsilon}=\tau_\pi \sqrt{\nabla_\mu \epsilon^\mu \nabla^\mu \epsilon}/\epsilon$, associated with the gradient of the energy density, were particularly relevant to assess the overall validity of the hydrodynamic description of the QGP formed in ultrarelativistic collisions.

\begin{figure}
\includegraphics[width=1\textwidth]{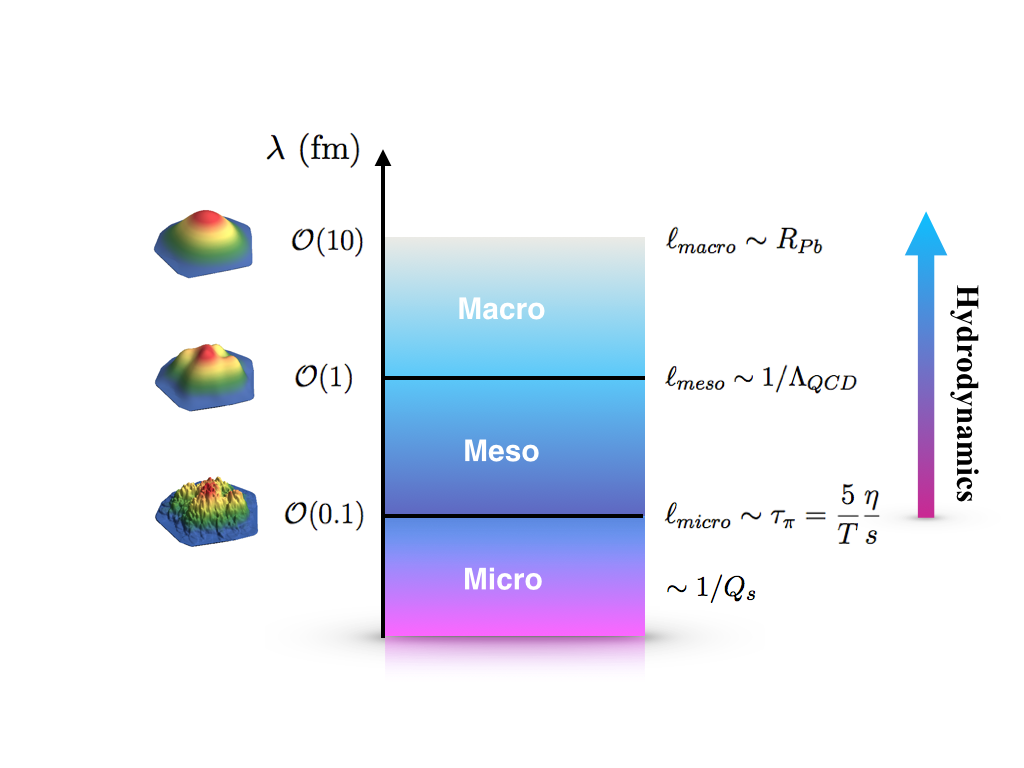}  
\caption{(Color online) Different length scales involved in the ultrarelativistic collisions of heavy nuclei separated according to the smoothing scale $\lambda$. The energy density snapshots on the right illustrate how the initial condition changes with the smoothing scale.} 
\label{fig:plotscales}
\end{figure}

The different length scales discussed above and the assumed region of applicability of IS hydrodynamics are summarized in Fig.\ \ref{fig:plotscales} where $\lambda$ is the \emph{smoothing scale} of the hydrodynamic simulations. We have also included snapshots of a typical initial energy density profile that show how the same event changes as the smoothing scale increases (this will be discussed in detail in the next section).  We also note that hydrodynamics cannot account for phenomena that occur at microscopic sub-nucleon scales at which the effects from coherent, color fields should not be neglected\footnote{We note that this transition from a color field dominated regime to hydrodynamics at large scales has also been extensively studied in the past in the context of jets \cite{CasalderreySolana:2004qm,Yarom:2007ni,Chesler:2007an,Gubser:2007ga,Noronha:2007xe,Noronha:2008un,Neufeld:2008fi,Neufeld:2008hs,Betz:2008wy,Betz:2008ka}. In fact, even in the case of a static and uniform medium, the small region near a fast moving colored parton is too influenced by the local color fields to display hydrodynamic behavior.}. Clearly, as the smoothing scale $\lambda$ increases, the local gradients $\sim 1/\lambda$ decrease (and so do the Knudsen numbers), and only the meso and macroscopic nuclear scale regimes are accessible in the hydrodynamic description.

The smoothing scale provides a natural way to characterize the different models for the initial conditions of hydrodynamics. For instance, models where the sources of fluctuations are related to the positions of the uncorrelated nucleons in the colliding nuclei, such as the Monte Carlo (MC) Glauber model \cite{Miller:2007ri,Alver:2008aq,Loizides:2014vua} and the MCKLN model \cite{Drescher:2006ca}, cannot properly resolve microscopic sub-nucleon scales (in the sense discussed above). On the other hand, the IP-Glasma model \cite{Schenke:2012wb} or the Dumitru-Nara BK model \cite{Dumitru:2012yr} depend on the physics at microscopic length scales of $\mathcal{O}(1/Q_s)$, which in turn produces large energy density fluctuations and structure even at very small scales. It is not clear at the moment how these very short wavelength fluctuations affect observable quantities related to hydrodynamical behavior such as the anisotropic flow coefficients.

In this paper we study the dependence of anisotropic flow coefficients on variations in the smoothing scale present in the initial conditions used in hydrodynamic simulations. The smoothing procedure is based on a Lagrangian description of hydrodynamics but it can also be employed to smooth out initial conditions that may later be used in (Eulerian) grid codes as well. The method can be used to smooth out initial conditions from both Glauber-like models as well as saturation based models such as IP-Glasma. In our approach, after going through this ``UV filter" defined by the smoothing scale $\lambda$, the initial condition only contains scales of size $\lambda$ and above: shorter scales are consistently smeared out. After this procedure, one can perform event by event viscous hydrodynamic simulations to investigate how smoothing out the short scales in the initial conditions may affect the hydrodynamic response of the fluid to the initial spatial inhomogeneities. 

We consider here MCKLN initial conditions \cite{Drescher:2006ca} for $\sqrt{s}=2.76$ TeV Pb+Pb collisions at the LHC and investigate how the anisotropic flow coefficients computed using the 2+1 viscous hydrodynamics code v-USPhydro \cite{Noronha-Hostler:2013ria,Noronha-Hostler:2013gga,Noronha-Hostler:2014dqa} are affected by changes in the smoothing scale. While the effects of different length scales present in the initial state have been investigated using different approaches \cite{Petersen:2010zt,ColemanSmith:2012ka,RihanHaque:2012wp,Bzdak:2013zma,Floerchinger:2013rya,Floerchinger:2013vua,Floerchinger:2013hza,Retinskaya:2013gca,Renk:2014jja,Konchakovski:2014fya}, an important novelty of our work is that the initial spatial eccentricities for nucleus-nucleus collisions (across centrality classes) do not change appreciably when the smoothing scale is varied in the mesoscopic regime ranging from 0.1 fm to 1 fm (a similar result was found in the context of the PHSD transport code \cite{Konchakovski:2014fya}). This result illustrates that, at least for the case of MCKLN initial conditions (and MCGlauber events), length scales below the mesoscopic scale ($\sim 1$ fm) given by the uncorrelated fluctuations in the position of the nucleons have a very small effect in the final azimuthal anisotropies in nucleus-nucleus collisions. Our results qualitatively agree with the conclusion of Ref.\ \cite{Floerchinger:2013rya} (obtained within mode-by-mode hydrodynamics) that the azimuthal anisotropies are much more sensitive to intermediate length scales (defined in the mesoscopic regime in our notation) than to scales deep in the microscopic sub-nucleon scale regime. Once the smoothing parameter enters the macroscopic nuclear scale regime, the eccentricities begin to change and if the process is continued even further all the structure is lost and nontrivial azimuthal anisotropies, such as triangular flow, would not be observed.

We also study how the variations in the smoothing scale affect the initial spatial eccentricities of top $1\%$ high multiplicity $\sqrt{s}=5.02$ TeV p+Pb collisions computed using the recently developed Trento code \cite{Moreland:2014oya}. We find that the eccentricities in p+Pb collisions are particularly sensitive to variations in the smoothing scale in the sub-nucleon regime. This sensitivity to the fluctuation scale in the mesoscopic regime is not seen in $65-70\%$ peripheral $\sqrt{s}=2.76$ TeV Pb+Pb collisions even though their multiplicity is comparable to high multiplicity p+Pb collisions \cite{Chatrchyan:2013nka}.

Also, in this paper we study how variations in the smoothing scale on an event by event basis affect $K_{n\theta}$ and the inverse Reynolds number, $Re^{-1}=\sqrt{\pi_{\mu\nu}\pi^{\mu\nu}}/P$ \cite{Denicol:2012cn} in Pb+Pb collisions. While $Re^{-1}$ remains generally small for most of the evolution\footnote{Our trivial choice for the initial conditions of the shear stress tensor at the initial time $\tau_0$, i.e, $\pi^{\mu\nu}(\tau_0,\mathbf{r})=0$ plays a big role to ensure this condition.} in the mesoscopic regime, $K_{n\theta}$ is significantly enhanced by the spatial inhomogeneities present in event by event simulations. However, given the robustness of the azimuthal flow coefficients with respect to variations of $\lambda$ in the mesoscopic regime in nucleus-nucleus collisions, hydrodynamic events with large local Knudsen numbers can be formally ``evolved" in $\lambda$ to reduce their local Knudsen number while keeping final hydrodynamic observables nearly unchanged.

This paper is organized as follows. In Section \ref{smoothingsection} we explain our smoothing procedure and show how it affects the initial energy density and spatial eccentricities in Pb+Pb collisions. In Section \ref{hydrosection} we give the details about our hydrodynamic simulations and show how variations in the smoothing parameter affect the Knudsen and inverse Reynolds number in an event by event simulation. We show in Section \ref{results} our results for the anisotropic flow coefficients in Pb+Pb collisions and make a comparison to LHC data. Our study on the dependence of the eccentricities with the smoothing scale in p+Pb collisions is done in Section \ref{pAsection}. We finish the paper in Section \ref{conclusions} with our conclusions and outlook. 

{\it Definitions}: We use a mostly minus metric $g_{\mu\nu} = (1,-1,-1,-\tau^2)$ expressed in Milne coordinates $x^\mu = (\tau,x,y,\varsigma)$ where 
\bea
\tau &=& \sqrt{t^2-z^2}\,, \\
\varsigma &=& \frac{1}{2}\ln \left(\frac{t+z}{t-z}\right)\,,
\eea
are the propertime and spacetime rapidity (expressed in terms of the standard Minkowski coordinates), respectively. Throughout this paper we use natural units where $\hbar=c=k_B=1$. Coordinates in the transverse plane are denoted as $\mathbf{r}=(x,y)$. 


\section{Smoothing out initial conditions for hydrodynamics}\label{smoothingsection}

Different assumptions regarding the underlying physics of the initial conditions for hydrodynamics in heavy ion collisions can lead to very different initial energy density profiles. For instance, MC Glauber-based models \cite{Alver:2008aq} produce much smoother (though of course still largely inhomogeneous) event by event energy density profiles than those found in IP-Glasma \cite{Schenke:2012wb}. It is of course of interest to check how hydrodynamic observables (such as the anisotropic flow) change with the spatial resolution present in the initial conditions. 

\subsection{Definition of the smoothing procedure}\label{smoothingsection1}

Initial condition models give (in one way or the other) the expectation value of the energy-momentum tensor of the matter, $T_{\mu\nu}(\tau_0,\mathbf{r})$, at an initial time $\tau_0$ for each event\footnote{In this paper we only discuss longitudinally boost invariant systems so no dependence on the spacetime rapidity is included. Clearly, our discussion can be generalized to include a nontrivial rapidity dependence as well, though in this case one may want to distinguish the smoothing scale in the transverse plane from that used in the rapidity direction.}. The idea here is to use some type of UV filter to systematically remove (on an event by event basis) from the initial $T_{\mu\nu}$ fluctuations of wavelength below a given smoothing scale $\lambda$. We shall denote the energy-momentum tensor that has gone through this $\lambda-$filter at the initial time by $T_{\mu\nu}(\tau_0,\mathbf{r};\lambda)$, i.e., 
\be
T_{\mu\nu}(\tau_0,\mathbf{r}) \xrightarrow{\text{\qquad$\lambda$-filter\qquad}} T_{\mu\nu}(\tau_0,\mathbf{r};\lambda)\,.
\ee 
To study how variations in the smoothing scale change the initial energy-momentum tensor one needs to solve a \emph{smoothing flow} functional equation 
\be
\frac{d}{d\lambda} T_{\mu\nu}(\tau_0,\mathbf{r};\lambda) = \mathcal{F}_{\mu\nu}\left[T_{\alpha\beta}(\tau_0,\mathbf{r};\lambda);\lambda\right],
\label{functionaleq}
\ee
where $\mathcal{F}_{\mu\nu}$ is a functional of the energy-momentum tensor. Clearly, the specific form of $\mathcal{F}_{\mu\nu}$ depends on the details regarding the implementation of the UV-filter. While the derivative in \eqref{functionaleq} is taken only with respect to the smoothing parameter, we expect that with the smoothing flow the spatial structures in the initial condition are smoothed out (or diffused) up to the minimal scale $\lambda$ and only the large scale properties of the energy-momentum tensor survive\footnote{One would be remiss to not point out here a loose resemblance with Ricci flow \cite{ricci}. In the latter, the metric of a manifold is evolved in a process formally analogous to heat diffusion to smooth out irregularities in a way that the knowledge of the geometric structure of the manifold can lead to useful global topological information. In our case involving hydrodynamics, the energy-momentum tensor is evolved in $\lambda$, removing small scale irregularities, while the global overall structure is preserved.}. Thus, given an initial $T_{\mu\nu}$ that displays structure in all the regimes displayed in Fig.\ \ref{fig:plotscales}, the smoothing flow\footnote{We would like to stress that the ``flow" in the smoothing procedure explained here has to do with variations in the underlying scale of the initial condition - no hydrodynamic evolution is done at that point and, thus, one should not identify this smoothing flow with the truly dynamical hydrodynamic flow.} procedure explained above can be used to systematically remove microscopic sub-nucleon scales and bring the system towards the meso or macroscopic nuclear scale regimes.  

A simple UV-filter that works as described above is the one already currently used in Lagrangian methods to solve the hydrodynamic equations, such as the Smoothed Particle Hydrodynamics (SPH) formalism \cite{originalSPH,SPHothers}, which has been successfully employed for more than a decade to study event by event simulations of numerical relativistic hydrodynamics in the context of heavy ion collisions within NeXSPheRIO \cite{Aguiar:2000hw,Osada:2001hw,Aguiar:2001ac,Socolowski:2004hw,Hama:2004rr,Andrade:2006yh,Andrade:2008xh,Takahashi:2009na,Gardim:2012yp} and also more recently in the v-USPhydro code \cite{Noronha-Hostler:2013ria,Noronha-Hostler:2013gga,Noronha-Hostler:2014dqa}. The SPH implementation used in v-USPhydro was discussed in detail in \cite{Noronha-Hostler:2013ria} and we shall review below only the points needed to define the UV-filter discussed above.

\begin{figure}
\begin{center}
\includegraphics[width=0.5\textwidth]{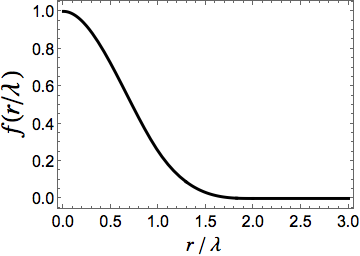}  
\caption{Function $f(|\mathbf{r}|/\lambda)$ used in the definition of the kernel in \eqref{definef}.} 
\label{fig:kernel}
\end{center}
\end{figure}

For simplicity, let us assume that the initial energy-momentum tensor is diagonal (i.e., there is no initial hydrodynamic flow or viscous contributions) such that we only need to consider the local energy density and pressure, which are assumed to be related via the equation of state (EOS). Of course, given the energy density the usual thermodynamical identities allow one to compute the local entropy density $s(\tau_0,\mathbf{r})$ and temperature $T(\tau_0,\mathbf{r})$ profiles. The main idea is that the initial energy density can always be reconstructed as follows: 
\be
\epsilon(\tau_0,\mathbf{r};\lambda) = \sum_{\alpha=1}^N \epsilon_\alpha(\tau_0)\,W\left(\frac{|\mathbf{r}-\mathbf{r}_\alpha|}{\lambda};\lambda\right),
\label{definefilter}
\ee
where $\epsilon_\alpha(\tau_0)$ denotes a discrete set of variables associated with different points in the transverse plane, $\left\{\mathbf{r}_{\alpha },\alpha =1,...,N\right\}$, where one places a piecewise continuous distribution function $W\left(|\mathbf{r}-\mathbf{r}_\alpha|/\lambda;\lambda\right)$. The kernel $W$ has finite support\footnote{The kernel is a delta sequence, i.e., $\lim_{\lambda\to 0}W\left(|\mathbf{r}|/\lambda;\lambda\right)=\delta(\mathbf{r})$ and is also normalized, i.e., $\int d^2\mathbf{r}\,W\left(|\mathbf{r}|/\lambda;\lambda\right)=1$.} given by $\lambda$ and, thus, it strictly vanishes for $|\mathbf{r}|/\lambda$ sufficiently larger than unit. The variables $\epsilon_\alpha(\tau_0)$ must be carefully chosen in order to better describe the structures in the initial condition for a given value of $\lambda$\footnote{See the discussion in \cite{deSouza:2015ena} based on the variational principle.}. In this work we use the cubic spline kernel employed in \cite{Noronha-Hostler:2013ria,Noronha-Hostler:2013gga,Noronha-Hostler:2014dqa} given by
\be
W\left(\frac{|\mathbf{r}|}{\lambda};\lambda\right) = \frac{10}{7\pi^2\lambda^2}\,f\left(\frac{|\mathbf{r}|}{\lambda}\right)\,,
\label{definef}
\ee
where
\begin{equation}
f\left(\xi\right)
=\left\{
     \begin{array}{lr}
       \mbox{if} \qquad \qquad  0\leq \xi<1   : &1-\frac{3}{2} \xi^2+\frac{3}{4}\xi^3,  \\
        \mbox{else if}  \qquad  1\leq \xi\leq2  :& \frac{1}{4}\left(2-\xi\right)^3, \\
        \mbox{else}   & 0\,.
     \end{array}
   \right.
\end{equation}
One can see in Fig.\ \ref{fig:kernel} that $f$ strictly vanishes when $|\mathbf{r}|/\lambda > 2$. Also, from \eqref{definefilter} one finds that (after filtering) the local gradient $\partial \epsilon\sim \mathcal{O}(1/\lambda)$, which implies that as $\lambda$ increases the energy density becomes smoother. However, it is important to notice that this filter somewhat preserves the relative peaks and valleys in the energy density. In fact, let $\mathbf{r}_\alpha$ be the position of a peak in the energy density. One can see that an increase in $\lambda$ does not change the location of the peak ($\epsilon_\alpha$ and $\mathbf{r}_\alpha$ remain the same) though its magnitude does decrease.   
In fact, for the implementation of the smoothing flow used in this paper one can see that
\be
\frac{d}{d\lambda}\epsilon(\tau_0,\mathbf{r};\lambda) = -\frac{2}{\lambda}\epsilon(\tau_0,\mathbf{r};\lambda)+\left(\frac{10}{7\pi^2\lambda^2}\right)\,\sum_{\alpha=1}^N \epsilon_\alpha(\tau_0)\,\frac{d}{d\lambda}f\left(\frac{|\mathbf{r}-\mathbf{r}_\alpha|}{\lambda}\right),
\ee
which illustrates the points discussed above. Also, even though the first term on the right hand side of the equation above ($-2 \epsilon/\lambda$) will be the same for any normalized kernel, the relative importance of the second term with respect to the first may depend on the properties of the kernel itself.




We remark that even though the discretization procedure used here was originally devised to be used in a Lagrangian approach to hydrodynamics\footnote{In fact, $\mathbf{r}_\alpha$ can be readily interpreted as the Lagrangian coordinates in a subsequent hydrodynamic evolution as done, for instance, in \cite{Noronha-Hostler:2013ria,Noronha-Hostler:2013gga,Noronha-Hostler:2014dqa}.}, the procedure explained in this section can be employed to smooth out initial energy density profiles that can be later used as input in Eulerian codes as well. 

\subsection{Smoothing out MCKLN events}

\begin{figure}
\centering
\includegraphics[width=1\textwidth]{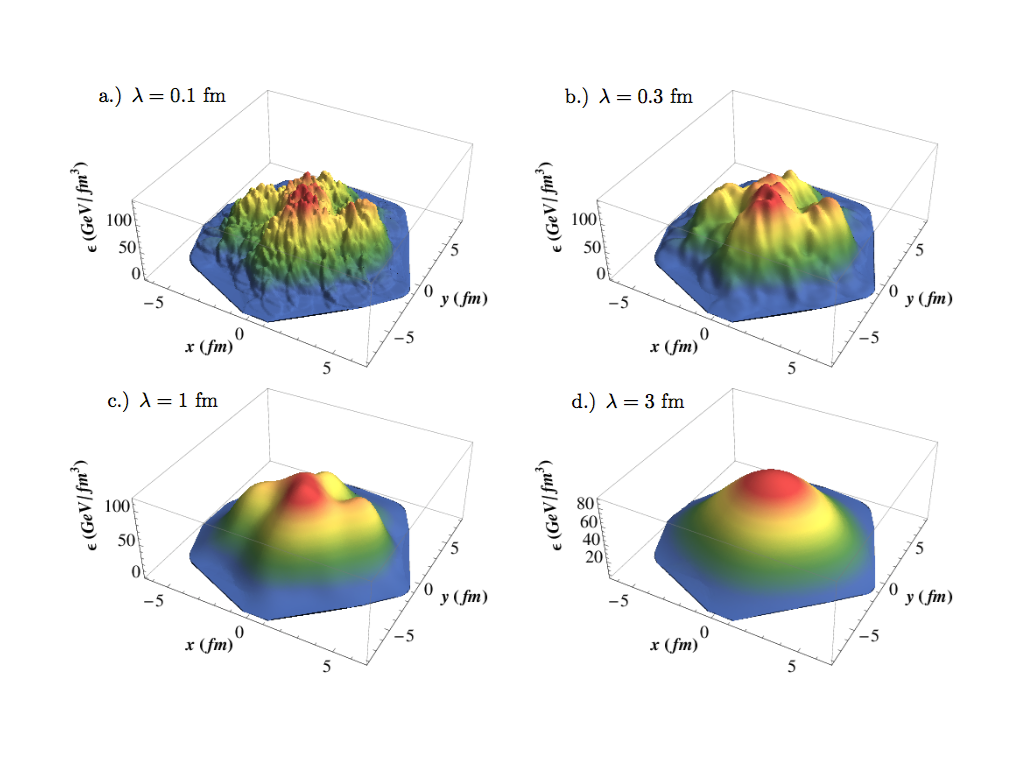} \\
\caption{(Color online) Effects of the smoothing procedure on the energy density profile of an MCKLN event in the $0-5\%$ centrality class of $\sqrt{s}=2.76$ TeV Pb+Pb collisions at the LHC. The following smoothing parameters were used: a.) $\lambda=0.1$ fm,  b.) $\lambda=0.3$ fm,  c.) $\lambda=1$ fm,  and d.) $\lambda=3$ fm.}
\label{fig:edenh}
\end{figure}

Here we apply the procedure discussed above to smooth out initial energy densities generated by the MCKLN code of \cite{Drescher:2006ca} for $\sqrt{s}=2.76$ TeV Pb+Pb collisions at the LHC. In Fig.\ \ref{fig:edenh} we show how the energy density of an event belonging to the $0-5\%$ centrality class changes if one increases the smoothing parameter from $\lambda=0.1$ to 3 fm. One can see that small inhomogeneities are progressively smeared out with increasing $\lambda$ until only the large scale energy density structure in panel d.), corresponding to the macroscopic nuclear scale regime mentioned in Fig.\ \ref{fig:plotscales}, is visible.

\subsubsection{Initial eccentricities}

\begin{figure}[ht!]
\centering
\begin{tabular}{cc}
\includegraphics[width=0.45\textwidth]{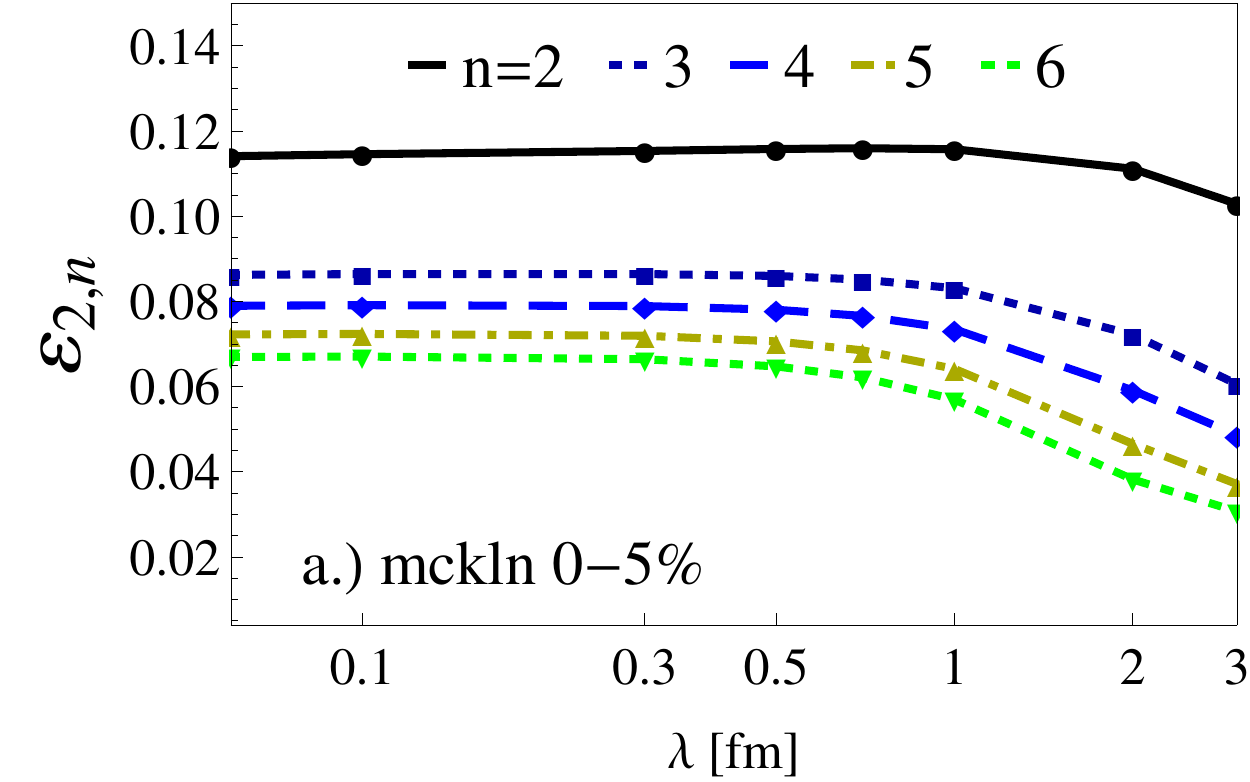} & \includegraphics[width=0.45\textwidth]{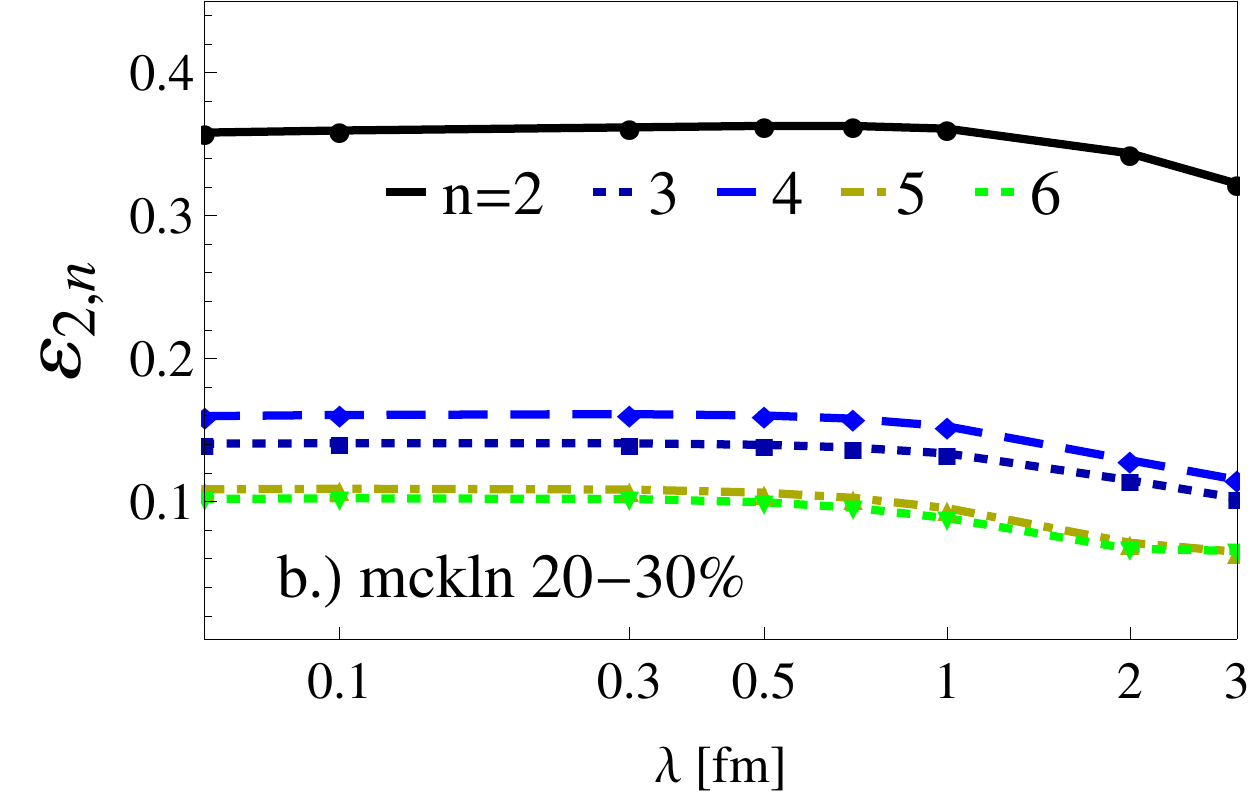}  \\
\end{tabular}
\caption{(Color online) Dependence of the initial eccentricities $\varepsilon_{2,n}\equiv \sqrt{\langle \hat\varepsilon_{2,n}^2 \rangle}$ (for $n=2-6$) on the smoothing parameter, $\lambda$, for $\sqrt{s}=2.76$ TeV Pb+Pb MCKLN events. The left plot was computed using events in the $0-5\%$ centrality class while on the right events in the $20-30\%$ centrality class were used (the average was performed over 150 events for each centrality class).}
\label{fig:ecc1}
\end{figure}

In this context, it is instructive to check how the initial spatial eccentricities vary with $\lambda$. The eccentricities computed event by event, $\hat\varepsilon_{m,n}$, acquire a $\lambda$ dependence due to their definition in terms of the initial energy density 
\be
\hat\varepsilon_{m,n}(\lambda)\,e^{i \hat\psi_{m,n}(\lambda)} = \frac{\int d^2\mathbf{r}\,\,r^m\, e^{in\phi}\,\epsilon(\tau_0,\mathbf{r};\lambda)}{\int d^2\mathbf{r}\,\,r^m\,\epsilon(\tau_0,\mathbf{r};\lambda)}\,,
\ee
where $r = |\mathbf{r}|$, $\phi = \tan^{-1}(y/x)$, and $\hat\psi_{m,n}$ are the initial angles. The dependence of $\hat\varepsilon_{2,n}$ ($n=2,\ldots,6$) with $\lambda$ can be seen in Fig.\ \ref{fig:ecc1} for two different centrality classes ($0-5\%$ on the left and $20-30\%$ on the right)\footnote{The initial energy densities before smoothing are defined on a square lattice of size $\lambda_0=0.06$ fm.}. In this plot we show $\varepsilon_{2,n} \equiv \sqrt{\langle \hat\varepsilon_{2,n}^2 \rangle}$ with the average being performed over 150 events in each centrality class. One can see that the eccentricities only change appreciably after one enters the macroscopic nuclear scale regime $\lambda > 1$ fm - this is the case for both centrality classes (we have also checked that this is the case for $0-1\%$ and $65-70\%$). This indicates that the scale that sets the upper limit of the mesoscopic scale is indeed of the order of $1/\Lambda_{QCD} \sim 1$ fm. However, note that higher order eccentricities, such as $\varepsilon_{2,6}$, begin to change for values of $\lambda$ smaller than $\varepsilon_{2,2}$ does. We remark that the robustness of the eccentricities with respect to variations of $\lambda$ found here are consistent with the previous general analysis performed in \cite{Bhalerao:2011bp}.

\begin{figure}[ht!]
\centering
\begin{tabular}{c c}
\includegraphics[width=0.45\textwidth]{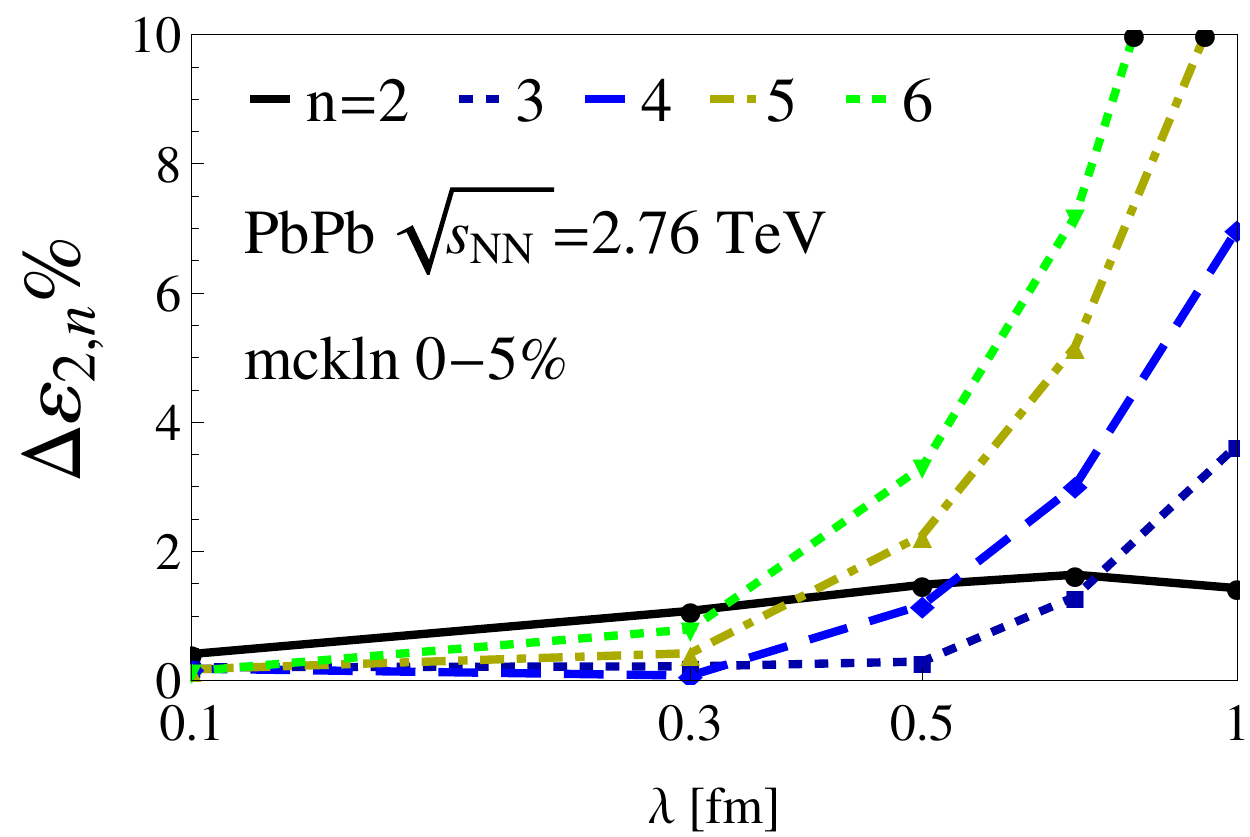} & \includegraphics[width=0.45\textwidth]{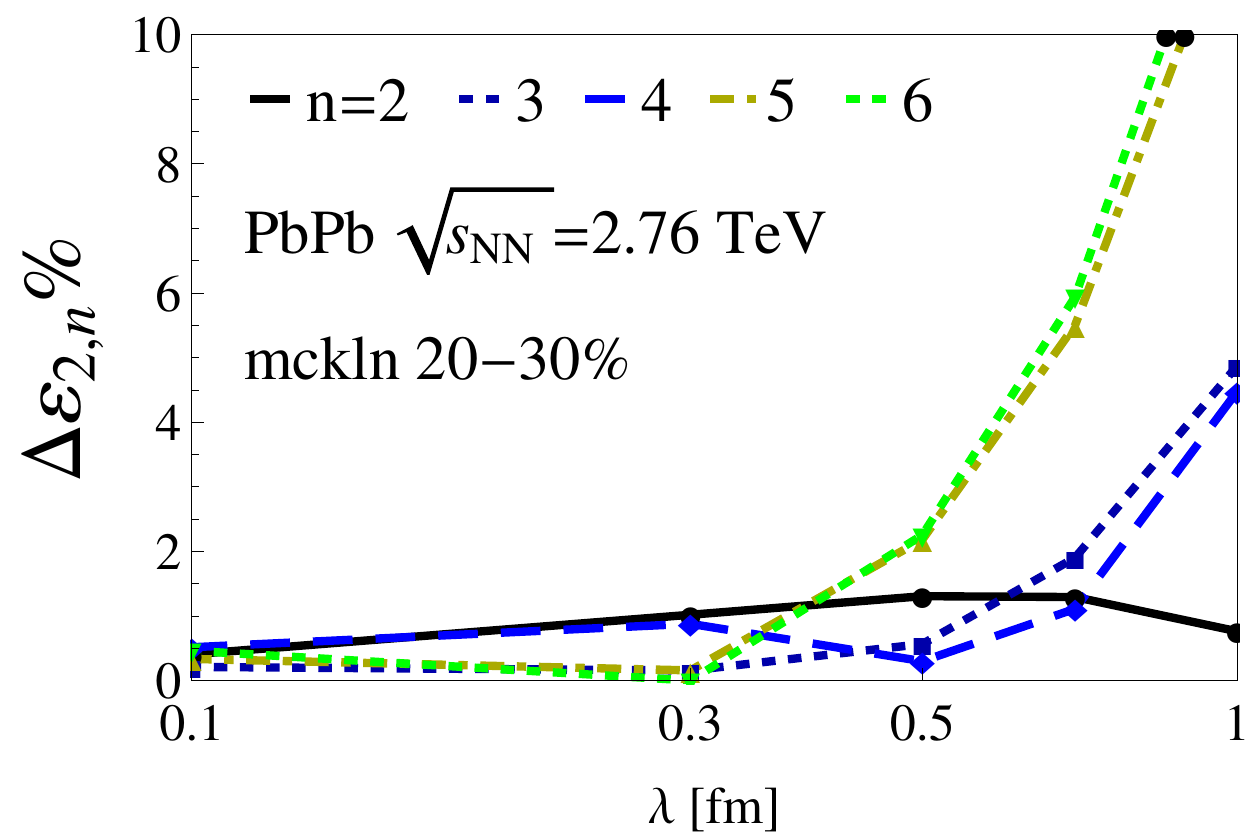} \\
\end{tabular}
\caption{(Color online) Relative variation of initial eccentricities $\Delta \varepsilon_{2,n}=100\left[\varepsilon_{2,n}(\lambda_0)-\varepsilon_{2,n}(\lambda)\right]/\varepsilon_{2,n}(\lambda_0)$ on the smoothing parameter, $\lambda$, for $\sqrt{s}=2.76$ TeV Pb+Pb MCKLN events. The left plot was computed using events in the $0-5\%$ centrality class while on the right events in the $20-30\%$ centrality class were used (the average was performed over 150 events for each centrality class).}
\label{fig:ecc3}
\end{figure}

The relative variation of the eccentricities $\Delta \varepsilon_{m,n}=100\left[\varepsilon_{m,n}(\lambda_0)-\varepsilon_{m,n}(\lambda)\right]/\varepsilon_{m,n}(\lambda_0)$ is displayed in Fig.\ \ref{fig:ecc3}, which shows that the lowest order eccentricities indeed change by less than $10\%$ when $\lambda$ varies from 0.1-1 fm (the higher order eccentricities are indeed more sensitive to smoothing, as expected). In summary, for MCKLN initial conditions, only when the smoothing scale is larger than 1 fm the essential spatial structures that contribute significantly to the eccentricities (and to the final azimuthal anisotropies) are washed out and the eccentricities themselves start to decrease significantly.
 
\begin{figure}[ht!]
\centering
\begin{tabular}{c c}
\includegraphics[width=0.45\textwidth]{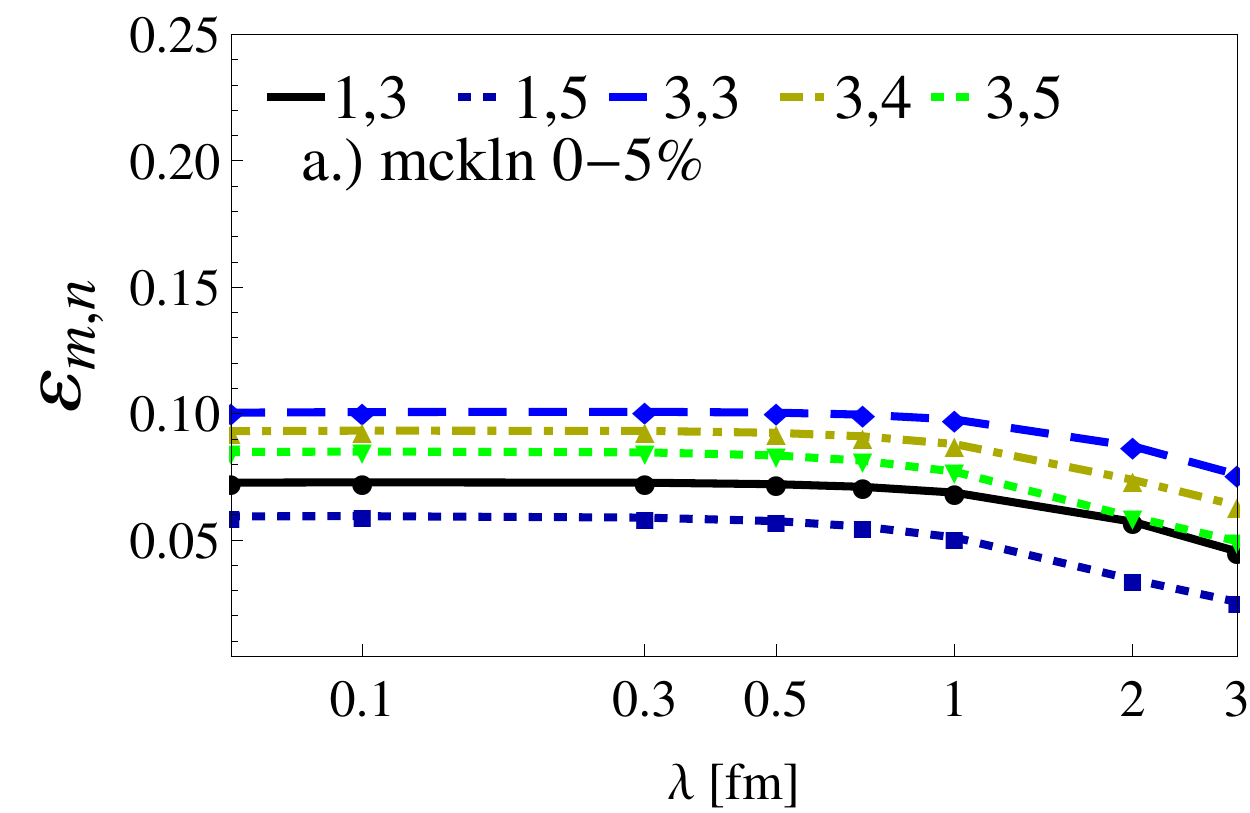} & \includegraphics[width=0.45\textwidth]{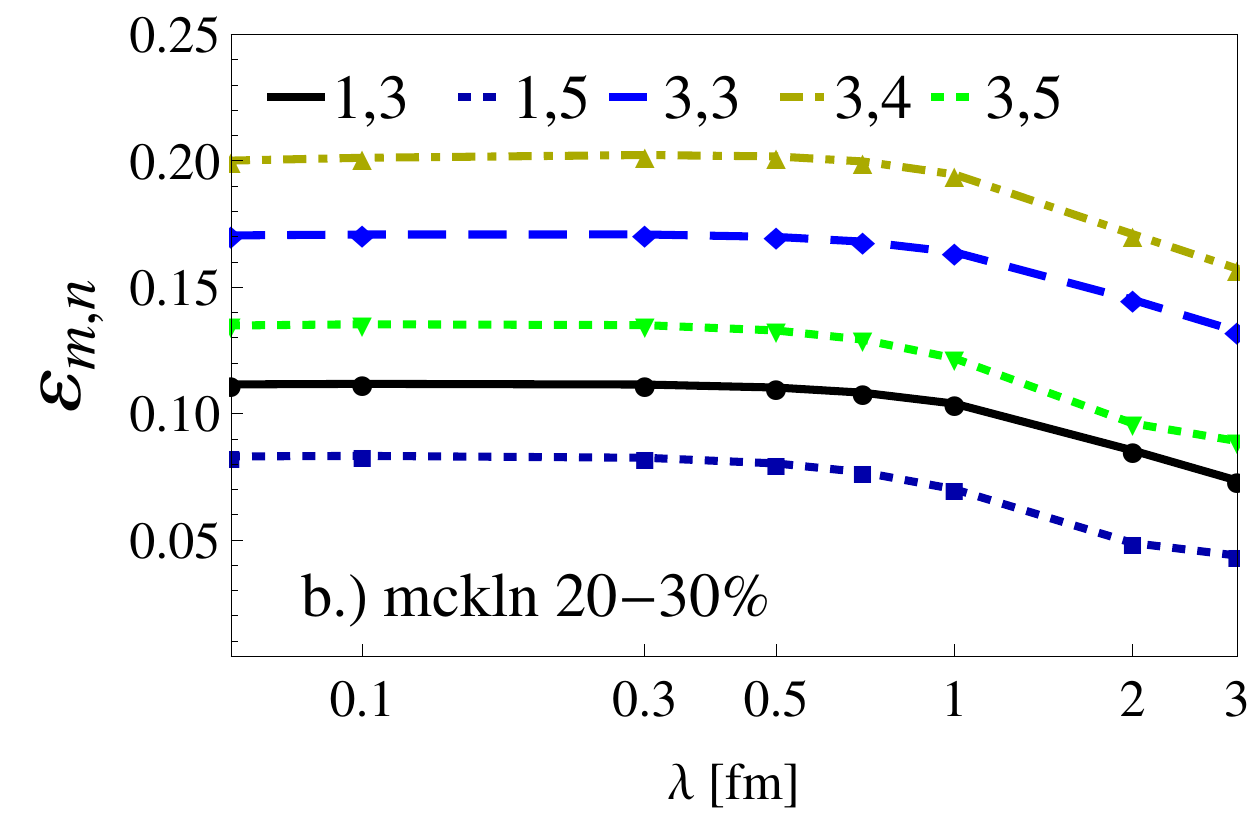} \\
\end{tabular}
\caption{(Color online) Dependence of the other initial eccentricities $\varepsilon_{m,n}=\sqrt{\langle \hat\varepsilon_{m,n}^2 \rangle}$ on the smoothing parameter, $\lambda$, for $\sqrt{s}=2.76$ TeV Pb+Pb MCKLN events. The left plot was computed using events in the $0-5\%$ centrality class while on the right events in the $20-30\%$ centrality class were used (the average was performed over 150 events for each centrality class).}
\label{fig:ecc2}
\end{figure}

The same conclusion can drawn about the other eccentricities shown in Fig.\ \ref{fig:ecc2}, which were previously shown to be relevant in the prediction of the final flow harmonics (see for instance, \cite{Gardim:2011xv,Gardim:2014tya}). Since the fluctuations in MCKLN have the same origin as in MCGlauber models (i.e., the random positions of the nucleons), we believe that the results found here concerning the dependence of eccentricities with the smoothing scale should also hold for MCGlauber initial conditions. It would be interesting to investigate how the eccentricities computed using the much finer IP-Glasma initial conditions \cite{Schenke:2012wb} vary with $\lambda$. One expects that, because of the presence of sub-nucleonic fluctuations in this type of model, its eccentricities should be more sensitive to the smoothing scale and significant changes may be found already for smaller values of $\lambda$ than those found here in MCKLN (i.e., the upper limit of the mesoscopic regime occurs at $\lambda < 1$ fm).

\section{Details about the hydrodynamic evolution and results for the Knudsen and inverse Reynolds numbers in nucleus-nucleus collisions}\label{hydrosection}

We used the v-USPhydro code \cite{Noronha-Hostler:2013ria,Noronha-Hostler:2013gga,Noronha-Hostler:2014dqa} to perform hydrodynamic simulations using as initial conditions the $\sqrt{s}=2.76$ TeV Pb+Pb MCKLN events in the $0-5\%$ and $20-30\%$ centrality classes discussed in the last section. The current version of v-USPhydro solves the energy-momentum conservation equations in Milne coordinates
\begin{equation}
\nabla_\mu T^{\mu\nu}=0 \Longrightarrow \frac{1}{\tau}\partial _{\mu }\left( \tau T^{\mu \nu }\right)
+\Gamma _{\beta \mu }^{\nu }T^{\beta \mu }=0 , \label{eqn:hydro}
\end{equation}%
where 
\begin{equation}
\Gamma _{\beta \mu }^{\nu }=\frac{1}{2}g^{\nu \sigma }\left( \partial
_{\mu }g_{\sigma \beta }+\partial _{\beta }g_{\sigma \mu }-\partial
_{\sigma }g_{\mu \beta }\right)
\end{equation}%
is the Christoffel symbol computed using the Milne metric $g_{\mu\nu} = (1,-1,-1,-\tau^2)$. Though bulk viscosity has been shown to be relevant in event by event hydrodynamic simulations \cite{Noronha-Hostler:2013gga,Noronha-Hostler:2014dqa,Gardim:2014tya,Rose:2014fba,Ryu:2015vwa}, for simplicity in this paper we focus solely on shear viscous effects and set the bulk viscosity to zero. Contributions from other 2nd order hydrodynamic transport coefficients besides $\tau_\pi$, such as those computed either in kinetic theory \cite{Denicol:2012cn} or in strongly coupled nonconformal holography \cite{Finazzo:2014cna}, are also neglected. 

The general expression for the energy-momentum tensor that includes shear viscosity effects is
\begin{equation}
T^{\mu \nu }=\epsilon\, u^{\nu}u^{\nu }-P \Delta ^{\mu
\nu }+\pi^{\mu\nu},
\label{emtensor}
\end{equation}%
where $\pi^{\mu\nu}$ is the shear stress tensor, $u^{\mu}$ is the fluid 4-velocity ($u_\mu u^\mu=1$), and the spatial projection operator is $%
\Delta _{\mu \nu }=g_{\mu \nu }-u_{\mu }u_{\nu }$.  We use the Landau
definition for the local rest frame, $u_{\nu }T^{\mu \nu }=\epsilon\,
u^{\mu }$. The shear stress tensor $\pi^{\mu\nu}$ obeys a simplified Israel-Stewart relaxation equation
\begin{equation} 
\tau_{\pi}\left(\Delta_{\mu\nu\alpha\beta}D\pi^{\alpha\beta}+\frac{4}{3}\pi_{\mu\nu}\theta\right)+\pi_{\mu\nu}=2\eta\sigma_{\mu\nu},
\label{shearreleq}
\end{equation}
where we have defined the tensor projector $\Delta_{\mu\nu\alpha\beta}=\frac{1}{2}\left[\Delta_{\mu\alpha}\Delta_{\nu\beta}+\Delta_{\mu\beta}\Delta_{\nu\alpha}-\frac{2}{3}\Delta_{\mu\nu}\Delta_{\alpha\beta}\right]$, the shear tensor $\sigma_{\mu\nu}=\Delta_{\mu\nu\alpha\beta}\nabla^\alpha u^\beta$, the comoving derivative $D=u^\mu \nabla_\mu$, and $\tau_{\pi} = 5\eta/(s T)$ is the shear viscosity relaxation time coefficient. In this paper we take a constant $\eta/s$ that is adjusted in each case to obtain a reasonable description of LHC data (see the discussion in Section \ref{resultshear}). This simplification overlooks the fact that $\eta/s$ may  strongly depend on the temperature when $T \sim 100-400$ MeV \cite{Hirano:2005wx,Csernai:2006zz,NoronhaHostler:2008ju,NoronhaHostler:2012ug}, which has been shown to affect elliptic flow \cite{Niemi:2011ix}. We leave an analysis of the effects of a temperature dependent $\eta/s$ (and $\zeta/s$) to a future publication.

The v-USPhydro code accurately\footnote{We note that v-USPhydro was shown \cite{Noronha-Hostler:2014dqa} to reproduce both the analytical and semi-analytical radially expanding solutions of Israel-Stewart hydrodynamics first developed in \cite{Marrochio:2013wla}.} solves the energy-momentum conservation equations \eqref{eqn:hydro} (in their boost invariant form) with $T^{\mu\nu}$ defined in \eqref{emtensor} together with the relaxation equation \eqref{shearreleq} for $\pi^{\mu\nu}$ using the SPH algorithm. The number of SPH particles in the simulations carried out here ranged from 40000 - 50000 depending on the centrality class. The SPH $h$ parameter for ideal hydrodynamics was set to be $h=0.1$ fm while we used $h=0.3$ fm in viscous simulations, as done in \cite{Noronha-Hostler:2013gga}. We used the lattice-based equation of state EOS S95n-v1 from \cite{Huovinen:2009yb} and an isothermal Cooper-Frye \cite{Cooper:1974mv} freezeout. The initial time for all the hydrodynamic simulations is $\tau_0 = 0.6$ fm. Particle decays are taken into account using an adapted version of the AZHYDRO code \cite{azhydro} with hadronic resonances with masses up to 1.7 GeV. Other technical details about our Lagrangian hydrodynamic solver can be found in \cite{Noronha-Hostler:2013gga}. 

\subsection{Effects of the smoothing procedure on Knudsen and inverse Reynolds numbers}

The behavior displayed by the eccentricities indicates that the final azimuthal anisotropies, computed within viscous hydrodynamics, may be rather insensitive to variations in $\lambda$ at least in the mesoscopic regime. Before we present our results for the flow anisotropies, below we first show how the smoothing procedure we implemented change the Knudsen and inverse Reynolds numbers on an event by event basis.

\begin{figure}[ht]
\centering
\begin{tabular}{cc}
\includegraphics[width=0.45\textwidth]{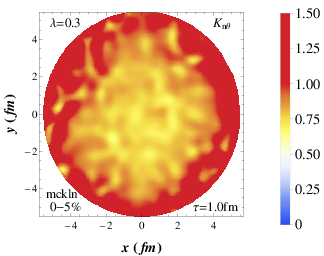}  & \includegraphics[width=0.45\textwidth]{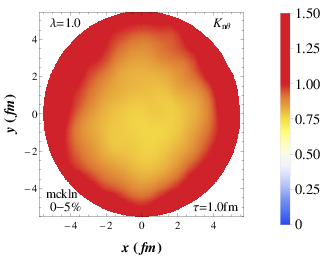}  \\
\includegraphics[width=0.45\textwidth]{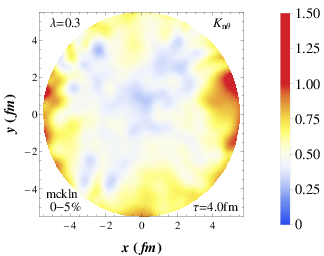}  & \includegraphics[width=0.45\textwidth]{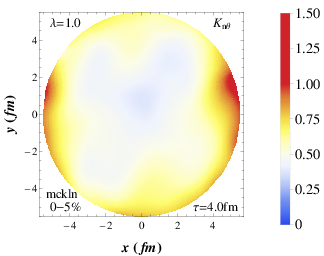}  \\
\end{tabular}
\caption{(Color online) Local Knudsen number, $K_{n\theta}=\tau_\pi\, \theta$, in the radius of $r<5.5$ fm for the same MCKLN event in the $0-5\%$ centrality class of $\sqrt{s}=2.76$ TeV Pb+Pb collisions at the LHC displayed in Fig.\ \ref{fig:edenh}. We use $\tau_\pi = 5\eta/(sT)$ and initial time $\tau_0=0.6$ fm. On the left we set $\lambda = 0.3$ fm (with $\eta/s=0.11$) and on the right $\lambda=1$ fm (with $\eta/s=0.1125$). The top panels were computed at $\tau=1$ fm while the panels at the bottom show $K_{n\theta}$ at $\tau=4$ fm.}
\label{fig:knudsen}
\end{figure}

We compute the Knudsen number $K_{n\theta} = \tau_\pi\, \theta$ \cite{Niemi:2014wta}, which is the product of the relevant microscopic scale, $\tau_\pi$, and a scalar that measures the spatial gradients of the flow, i.e, the expansion rate $\theta$. We took the minimal value of $\tau_\pi$ in our viscous simulations, $\sim 0.3$ fm, as as lower bound on the value of the smoothing parameter $\lambda$ in viscous simulations to ensure that we are not evolving hydrodynamically truly microscopic sub-nucleon scales that are beyond the reach of Israel-Stewart hydrodynamics\footnote{If one were to use the shear relaxation time computed at strong coupling in \cite{Finazzo:2014cna} which is roughly $\sim 0.2/T$, this lower bound would be approximately a factor of two smaller. This would essentially bring down the local $K_{n\theta}$ in Fig.\ \ref{fig:knudsen} by a factor of two. Therefore, one can see that the underlying assumptions about the microscopic nature of the fluid, i.e., weak (based on kinetic theory) or strong coupling, can shift up or down the ballpark estimate for the microscopic scale relevant for hydrodynamics.}. 

In Fig.\ \ref{fig:knudsen} we show $K_{n\theta}=\tau_\pi\, \theta$ within the radius of $r<5.5$ fm for the same MCKLN event in Fig.\ \ref{fig:edenh} that belongs to the $0-5\%$ centrality class of $\sqrt{s}=2.76$ TeV Pb+Pb collisions at the LHC. On the left we set the smoothing scale of the initial conditions to be $\lambda = 0.3$ fm (with $\eta/s=0.11$) and on the right $\lambda=1$ fm (where $\eta/s=0.1125$). The top panels were computed at $\tau=1$ fm (i.e., 0.4 fm after the initial time) while the panels at the bottom show $K_{n\theta}$ at $\tau=4$ fm. At early times (top panels), most of the matter is within the $5.5$ fm radius we plotted and most of the regions where $K_{n\theta} >1$ form the edge of the system. However, note that even if one changes $\lambda$ by nearly an order of magnitude, the Knudsen number in the inner region is never small though most of details of the spatial structure are washed out when going from $\lambda=0.3$ to 1 fm. In fact, for $\lambda=1$ fm a disk of 2 fm radius centered at the origin would correspond to a nearly uniform $K_{n\theta} \sim 0.75$ while when $\lambda=0.3$ fm the same 2 fm radius disk would still contain some subregions where $K_{n\theta} = 1$. At later times (bottom panels), the system has had enough time to expand and the region where $K_{n\theta}>1$ has been pushed outside of the plotted disk, though the inner region still shows moderately large values around 0.5. These results show that the local Knudsen number decreases when one increases the smoothing scale of the initial condition (as expected), though the (lowest order) spatial eccentricities shown in Fig.\ \ref{fig:ecc1} can remain nearly unchanged.  

\begin{figure}[ht]
\centering
\begin{tabular}{cc}
\includegraphics[width=0.45\textwidth]{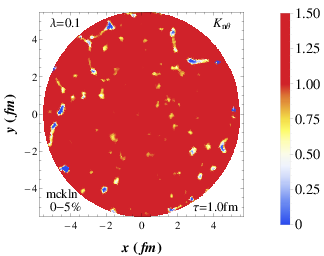}  & \includegraphics[width=0.45\textwidth]{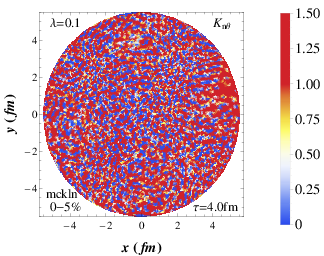}  \end{tabular}
\caption{(Color online) Local Knudsen number, $K_{n\theta}=\tau_\pi\, \theta$, in the radius of $r<5.5$ fm for the same MCKLN event in the $0-5\%$ centrality class of $\sqrt{s}=2.76$ TeV Pb+Pb collisions at the LHC displayed in Fig.\ \ref{fig:edenh}. We use again $\tau_\pi = 5\eta/(sT)$ and initial time $\tau_0=0.6$ fm though now $\lambda=0.1$ fm and $\eta/s=0.03$. On the left show $K_{n\theta}$ at $\tau=1$ fm while on the right $\tau=4$ fm.}
\label{fig:knudsenbad}
\end{figure}

To illustrate the issue with the applicability of hydrodynamics in event by event simulations containing initial state fluctuations of very short wavelength, we show in Fig.\ \ref{fig:knudsenbad} the same event as before but now we take $\lambda=0.1$ fm. Even with the very small $\eta/s=0.03$ used in this calculation, one can see that the Knudsen number remains large and highly inhomogeneous throughout the hydrodynamic evolution. A comparison between Figs.\ \ref{fig:knudsenbad} and \ref{fig:knudsen} show that using the minimal value of $\tau_\pi\sim 0.3$ fm in our simulations as a lower bound on $\lambda$ gives much more sensible Knudsen number profiles where the applicability of hydrodynamics can be at least suggested - the same cannot be said about the case where $\lambda=0.1$ fm. 

\begin{figure}[ht]
\centering
\begin{tabular}{cc}
\includegraphics[width=0.45\textwidth]{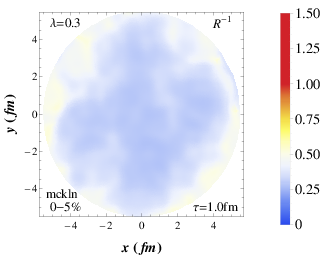}  & \includegraphics[width=0.45\textwidth]{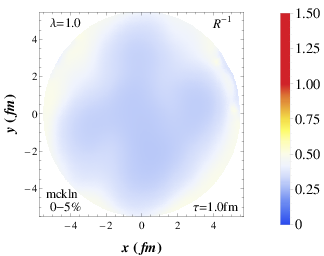}  \\ \includegraphics[width=0.45\textwidth]{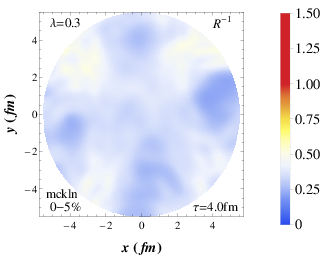}  & \includegraphics[width=0.45\textwidth]{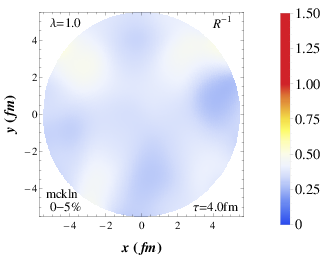}  \\
\end{tabular}
\caption{(Color online) Local inverse Reynolds number, $Re^{-1}=\sqrt{\pi_{\mu\nu}\pi^{\mu\nu}}/P$, in the radius of $r<5.5$ fm for the same MCKLN event in the $0-5\%$ centrality class of $\sqrt{s}=2.76$ TeV Pb+Pb collisions at the LHC displayed in Fig.\ \ref{fig:edenh}. We use $\tau_\pi = 5\eta/(sT)$ and initial time $\tau_0=0.6$ fm at which $\pi^{\mu\nu}=0$. On the left we set the smoothing scale to be $\lambda = 0.3$ fm (with $\eta/s=0.11$) and on the right $\lambda=1$ fm (with $\eta/s=0.1125$). The top panels were computed at $\tau=1$ fm while the panels at the bottom show $Re^{-1}$ at $\tau=4$ fm.}
\label{fig:reynolds}
\end{figure}

In Fig.\ \ref{fig:reynolds} we show our results for the inverse Reynolds number, $Re^{-1}=\sqrt{\pi_{\mu\nu}\pi^{\mu\nu}}/P$, for the same event. Once more, $\lambda = 0.3$ fm on the left while on the right $\lambda=1$ fm. The top panels correspond to $\tau=1$ fm and the bottom panels  show $Re^{-1}$ at $\tau=4$ fm. Given that $\pi^{\mu\nu}(\tau_0,\mathbf{r})=0$, one can see that $Re^{-1}$ is small in the 5.5 fm radius disk we plotted for both early and late times and different values of $\lambda$. It would be interesting to investigate the case where $\pi^{\mu\nu}$ is not set to zero at $\tau_0$ to see how its initial inhomogeneities and further time evolution are affected by the smoothing procedure considered in this paper. 

\section{Results for the anisotropic flow coefficients in nucleus-nucleus collisions}\label{results}

We have seen in the previous sections that variations of the smoothing scale $\lambda$ by nearly an order of magnitude within the mesoscopic regime (see Fig.\ \ref{fig:plotscales}) do not change significantly the initial spatial eccentricities in nucleus-nucleus collisions, though the local Knudsen number does decrease with increasing $\lambda$. In this section we study how smoothing out the initial conditions affect the anisotropic flow coefficients in Pb+Pb collisions. We use the event plane method \cite{Poskanzer:1998yz} to compute the anisotropic flow coefficients $v_n$ (see Appendix C of \cite{Noronha-Hostler:2013gga} for details) of the 150 individual events in each centrality class.  

\subsection{Flow coefficients within ideal hydrodynamics}

Here we study how changes in $\lambda$ affect the flow harmonics computed in the inviscid limit, i.e., $\eta/s=0$. After performing the smoothing procedure on the initial energy density, the particle spectrum (before integration over the azimuthal angle) acquires a dependence on the smoothing scale $\lambda$, which is later passed over to the final $v_n$'s obtained after freezeout+decay. We have set the freezeout temperature to be $T_{FO}=130$ MeV in the ideal hydrodynamic calculations. 

\begin{figure}[ht!]
\centering
\includegraphics[width=0.6\textwidth]{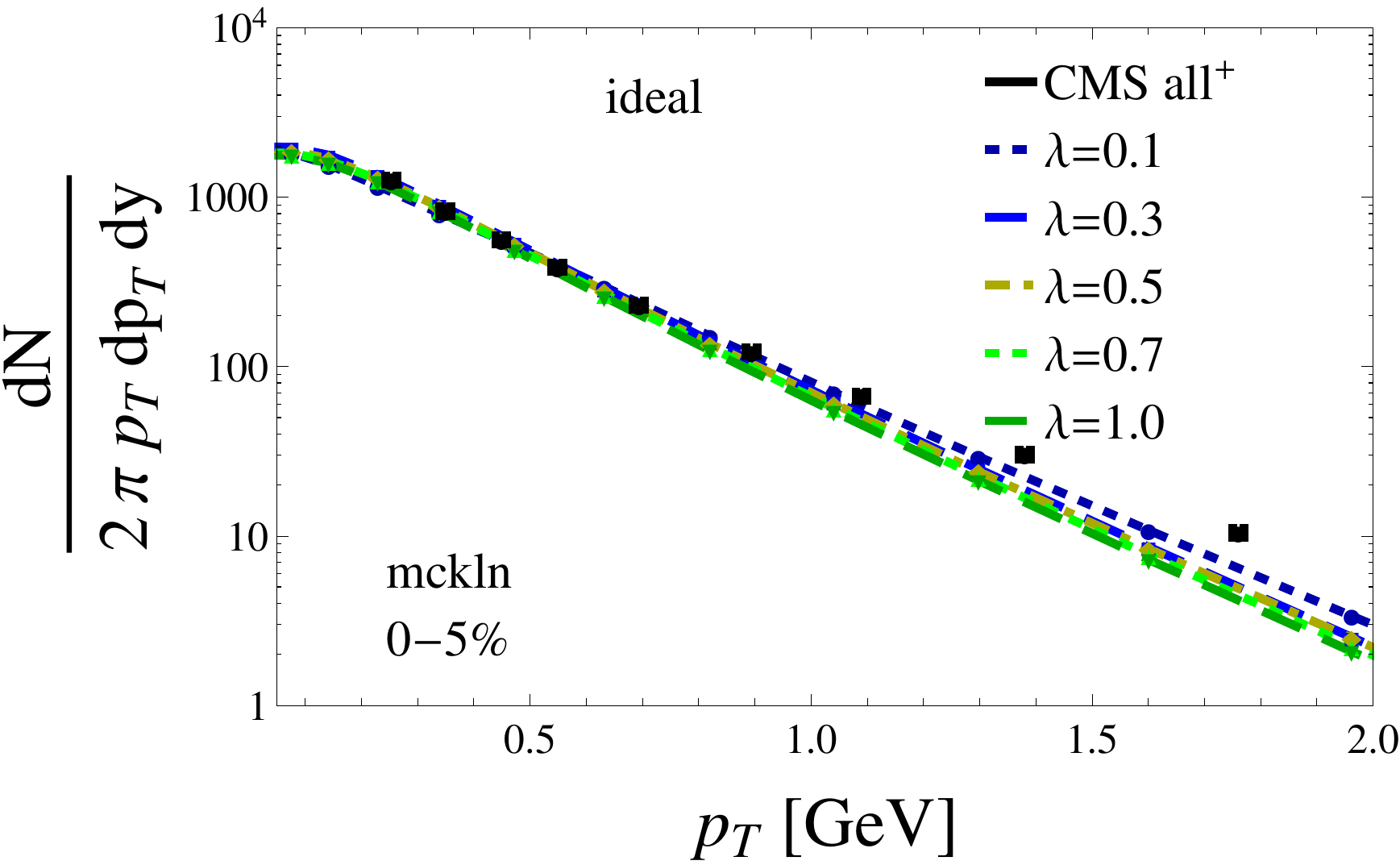}
\caption{(Color online) $dN/(2\pi p_Tdp_Tdy)$ of all charged particles in the $0-5\%$ centrality class of $\sqrt{s}=2.76$ TeV Pb+Pb collisions at the LHC computed using ideal hydrodynamics (averaged over 150 events) for $\lambda=0.1-1.0$ fm. Data points correspond to CMS data \cite{Chatrchyan:2012ta,Chatrchyan:2013kba}. }
\label{fig:speclam}
\end{figure}

In Fig.\ \ref{fig:speclam} we show how the spectrum $dN/(2\pi p_Tdp_Tdy)$ of all charged particles for the $0-5\%$ centrality class, computed using ideal hydrodynamics, is affected when $\lambda$ varies from $0.1$ to 1 fm. CMS data \cite{Chatrchyan:2012ta,Chatrchyan:2013kba} is included in this plot to give an idea about the magnitude of the effect. While at sufficiently low $p_T < 0.5$ GeV the spectrum can be made robust with respect to variations in $\lambda$ (by appropriately choosing the overall constant in the initial energy density), as one increases $\lambda$ the high $p_T$ part is depleted and the spectrum gets softer. This is expected since the short wavelength fluctuations present in event by event simulations that generally contribute to the high $p_T$ part of the spectrum \cite{Hama:2004rr,Andrade:2008xh} are systematically removed when going from $\lambda=0.1$ to 1 fm, which in turn decreases the average transverse momentum. 

\begin{figure}[ht!]
\centering
\begin{tabular}{ c c}
\includegraphics[width=0.4\textwidth]{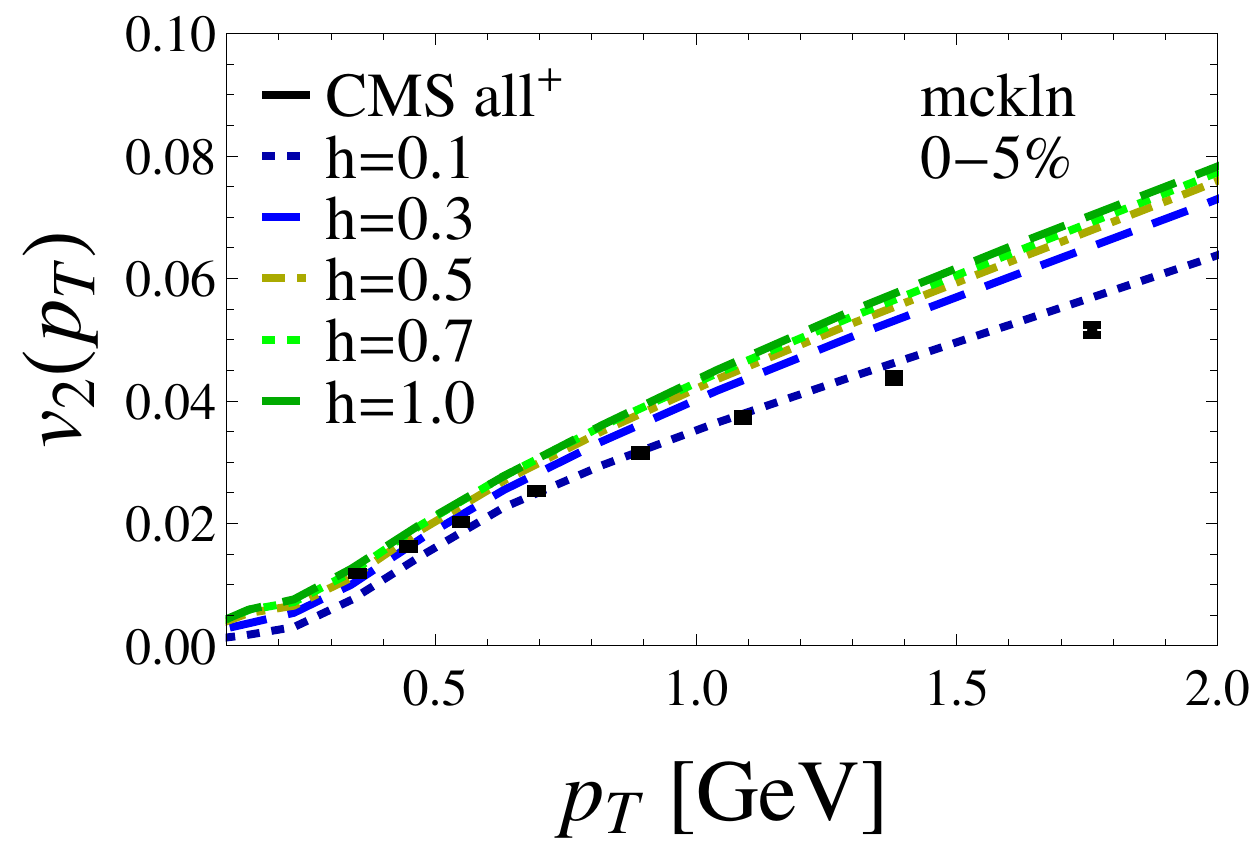} & \includegraphics[width=0.4\textwidth]{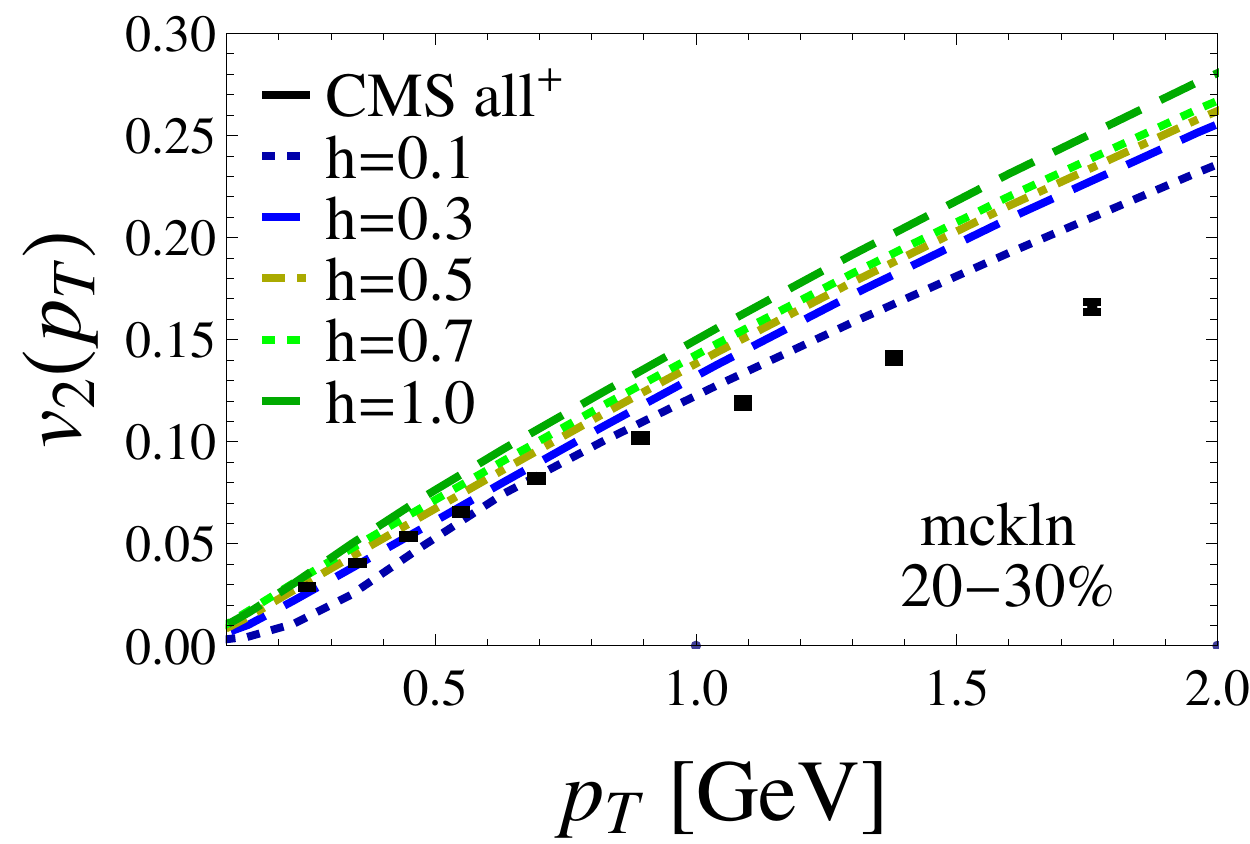}  \\
\includegraphics[width=0.4\textwidth]{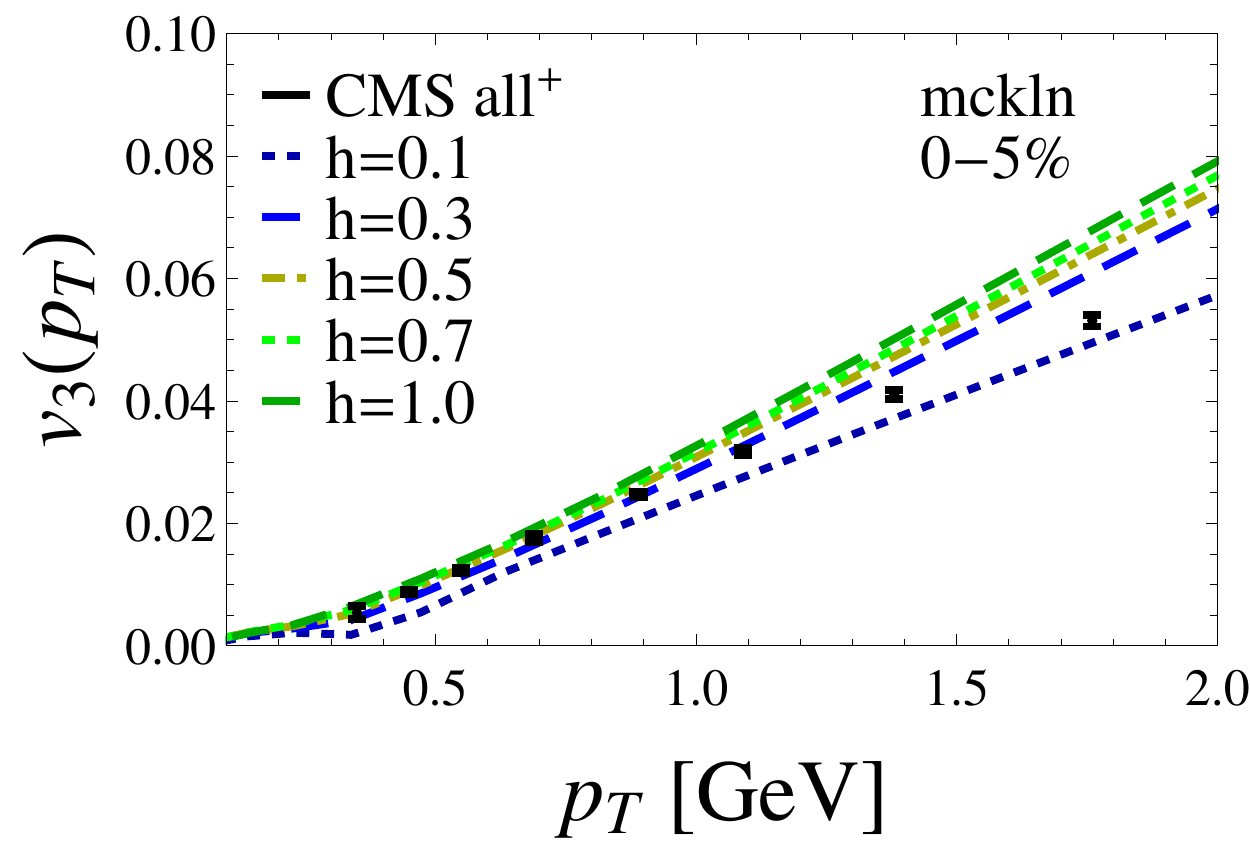} & \includegraphics[width=0.4\textwidth]{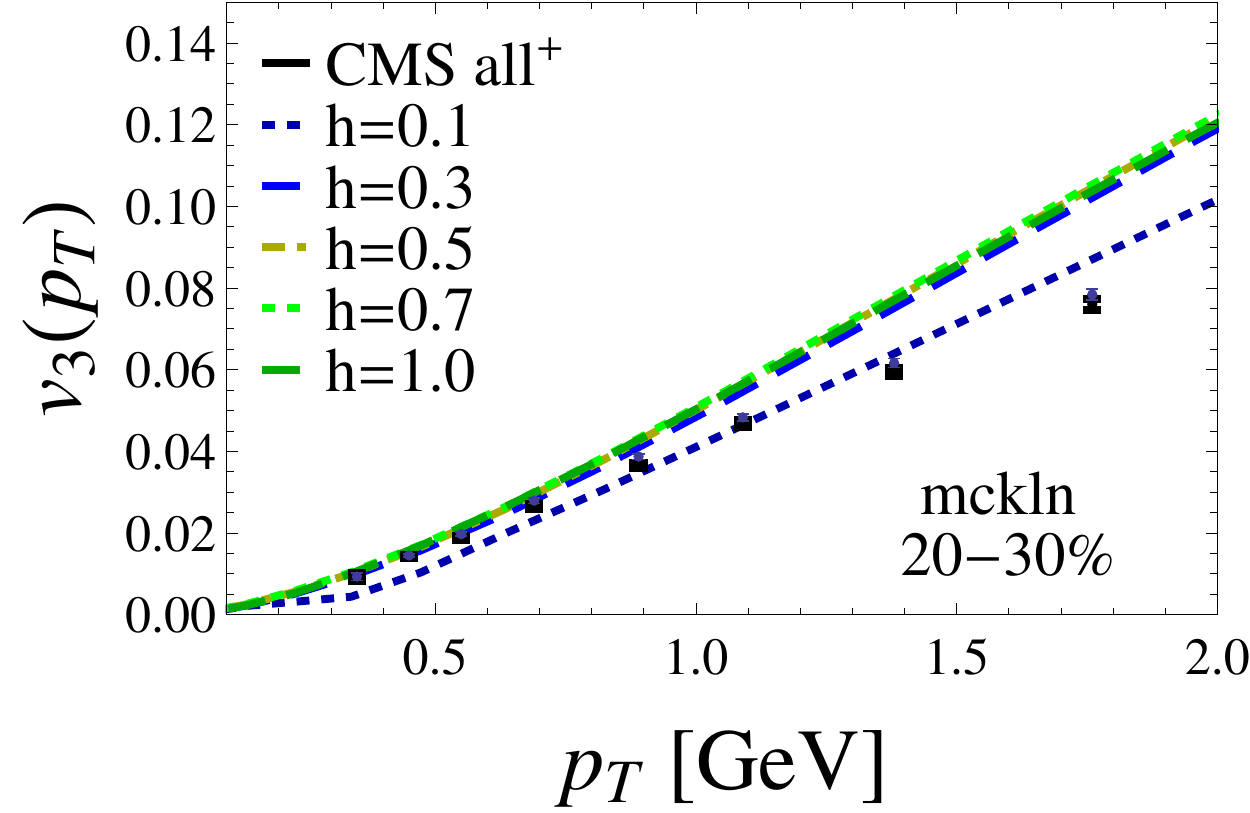}   \\
\includegraphics[width=0.4\textwidth]{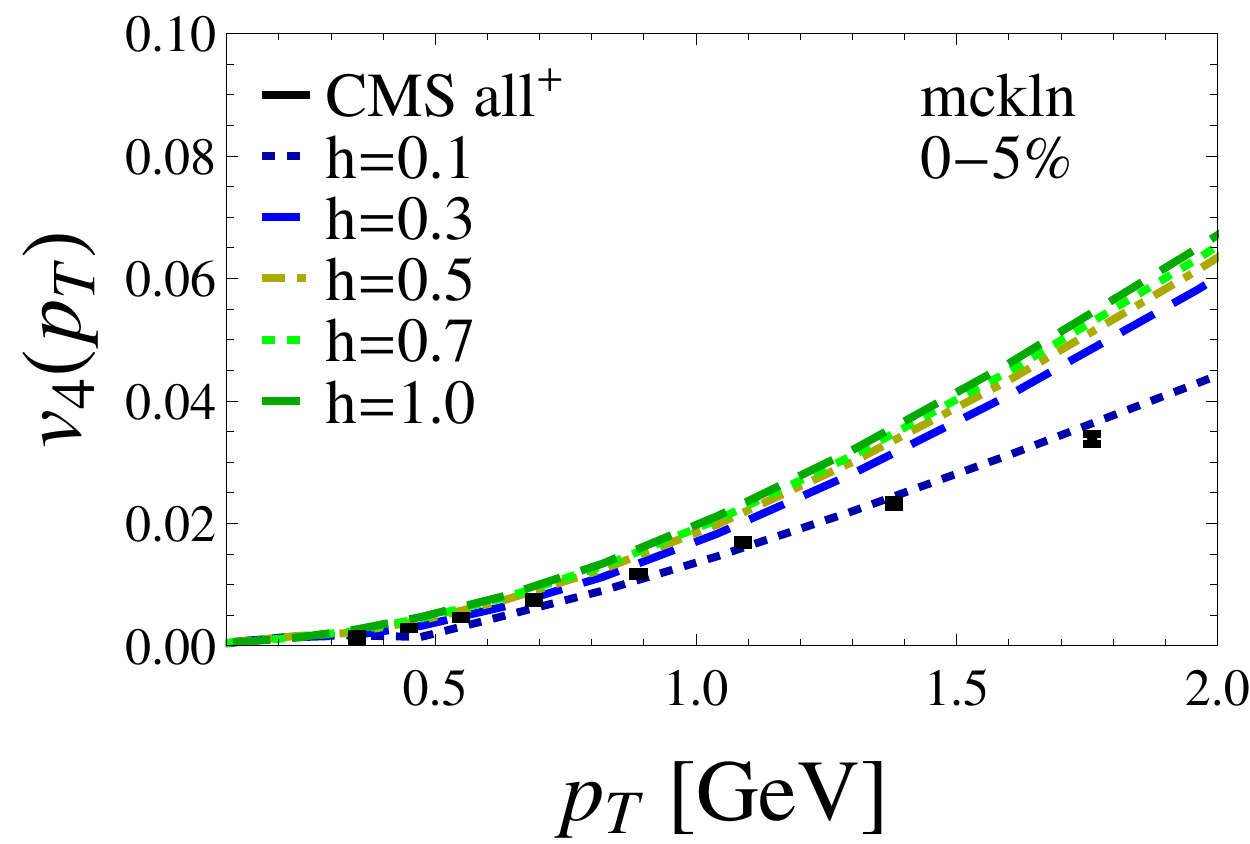} & \includegraphics[width=0.4\textwidth]{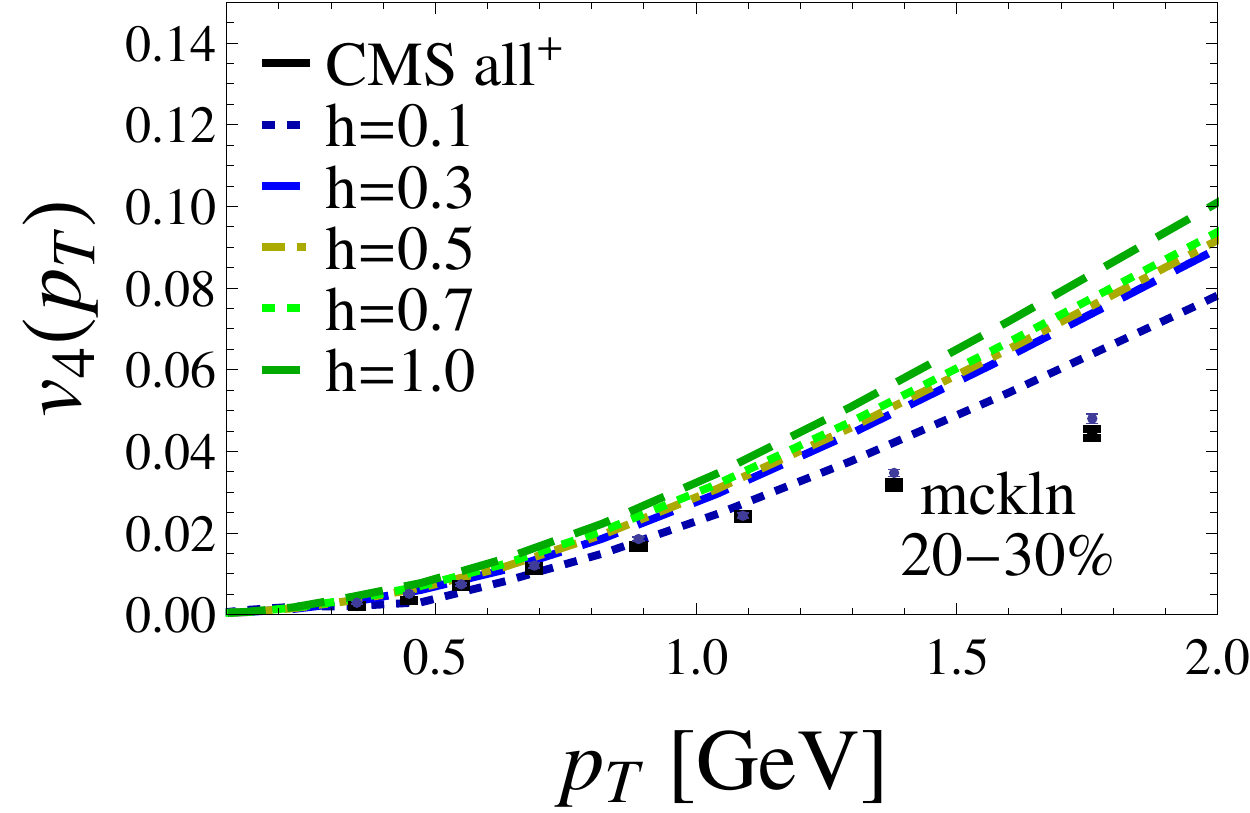}  \\
\includegraphics[width=0.4\textwidth]{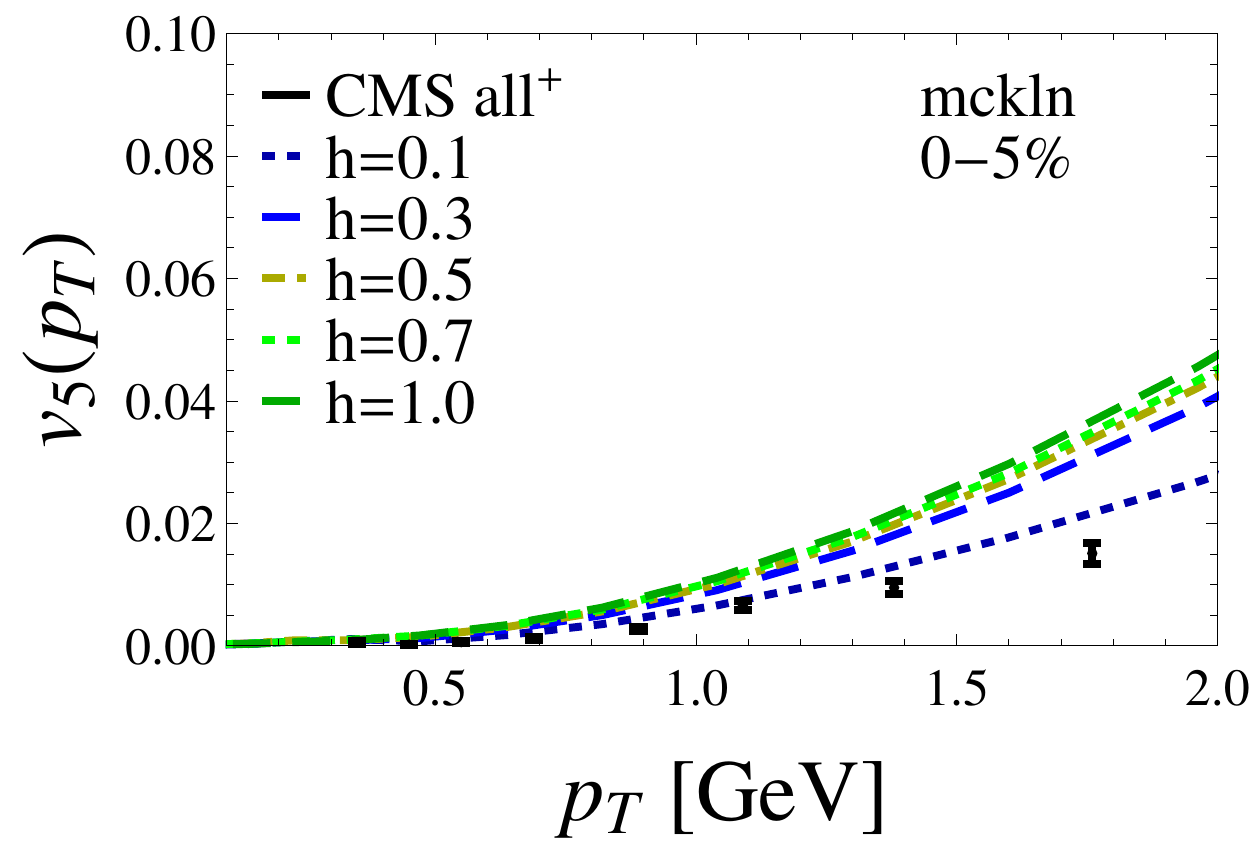} & \includegraphics[width=0.4\textwidth]{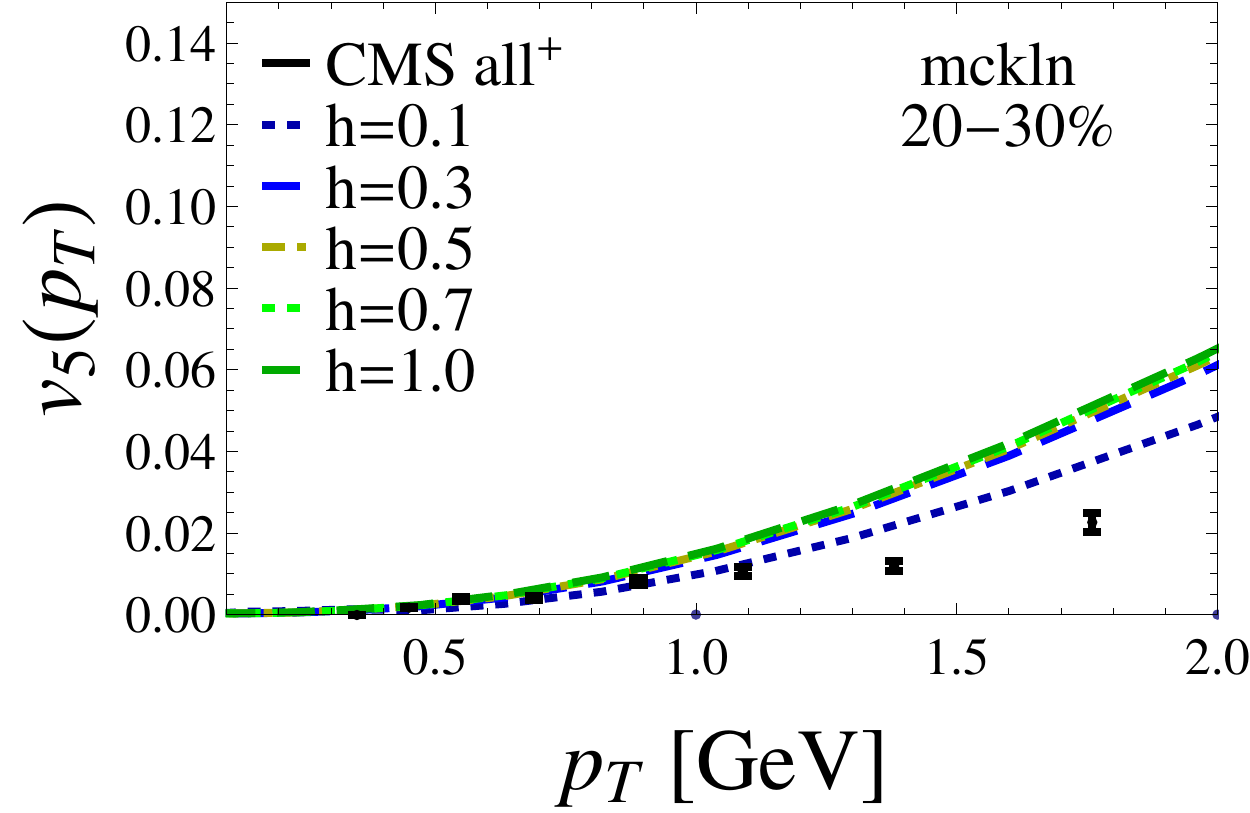} \\
\end{tabular}
\caption{(Color online) Convergence of the differential flow harmonics of all charged particles $v_n(p_T)\equiv \sqrt{\langle v_n^2(p_T)\rangle}$ (computed within ideal hydrodynamics and averaged over 150 events in each centrality class) as one increases the smoothing parameter $\lambda$ from $0.1$ to 1 fm. The left panels correspond to the $0-5\%$ centrality class while the right panels show the $20-30\%$ centrality results for $\sqrt{s}=2.76$ TeV Pb+Pb collisions at the LHC. The data points correspond to CMS data \cite{Chatrchyan:2012ta,Chatrchyan:2013kba}.}
\label{fig:vnpth}
\end{figure}

In Fig.\ \ref{fig:vnpth} we show how the differential flow harmonics of all charged particles $v_n(p_T)\equiv \sqrt{\langle v_n^2(p_T)\rangle}$ (computed within ideal hydrodynamics and averaged over 150 events in each centrality class) change as one increases the smoothing parameter $\lambda$ from $0.1$ to 1 fm. The left panels correspond to the $0-5\%$ centrality class while the right panels show the $20-30\%$ centrality results for $\sqrt{s}=2.76$ TeV Pb+Pb collisions at the LHC. Once more, the data points here correspond to CMS data \cite{Chatrchyan:2012ta,Chatrchyan:2013kba}, which are only included at this point to give an idea of the magnitude of the generated anisotropic flow. There are a few salient features encoded in these plots. First, in accordance with the results for the spectrum discussed before, one sees that the anisotropic flow coefficients are enhanced when short wavelength fluctuations are smoothed out\footnote{Ref.\ \cite{Andrade:2008xh} observed the same behavior in their calculations of elliptic flow.}. In our case, the addition of short wavelength fluctuations bring the $v_n$'s down because their contribution to the overall anisotropy is not sufficient to overcome the increase in the particle spectrum in Fig.\ \ref{fig:speclam} for small values of $\lambda$. Note also that ideal hydrodynamics with $\lambda=0.1$ fm is already very close to the data ($v_3$ is actually below the data) especially for $0-5\%$ centrality. Another interesting feature displayed in Fig.\ \ref{fig:vnpth} is that $v_n(p_T)$ changes less when $\lambda$ goes from 0.3 to 1 fm than it does when $\lambda$ varies between $0.1$ and 0.3 fm. Moreover, we found that odd harmonics converge faster than their even counterparts. 

\begin{figure}[ht]
\centering
\begin{tabular}{cc}
\includegraphics[width=0.45\textwidth]{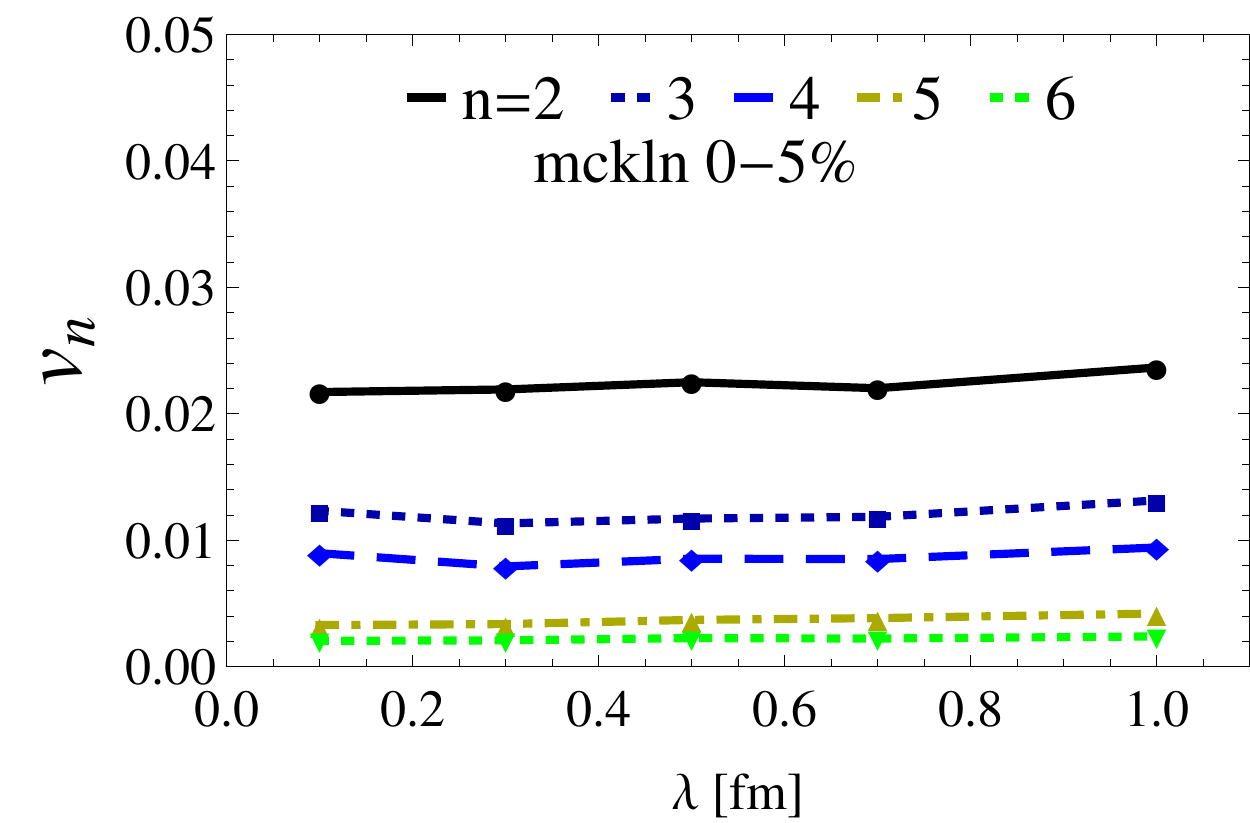} & \includegraphics[width=0.45\textwidth]{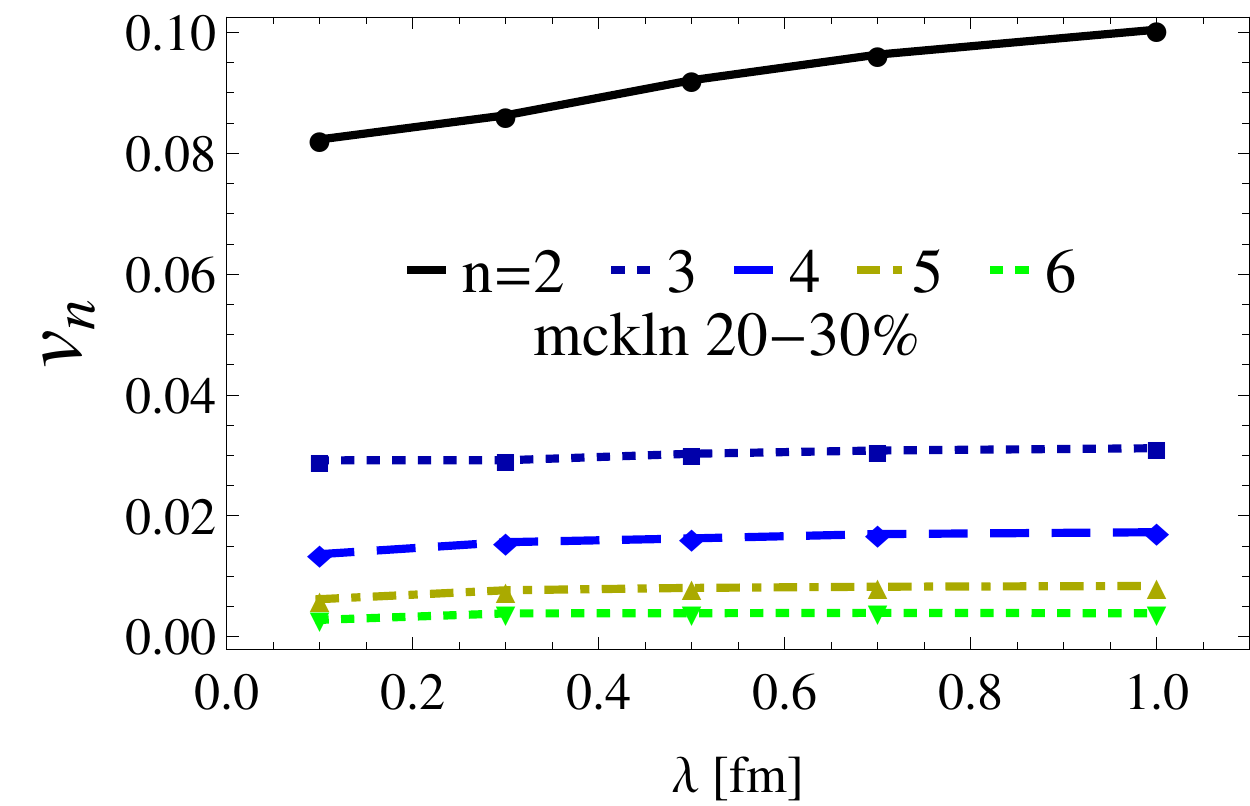}  
\end{tabular}
\caption{(Color online) Dependence of the integrated flow harmonics $v_n\equiv \sqrt{\langle v_n^2\rangle}$ (computed within ideal hydrodynamics and averaged over 150 events in each centrality class) on the smoothing scale $\lambda$. The left panel corresponds to the $0-5\%$ centrality class while the right panel shows the $20-30\%$ centrality results for $\sqrt{s}=2.76$ TeV Pb+Pb collisions at the LHC.}
\label{fig:int}
\end{figure}

In Fig.\ \ref{fig:int} we show the corresponding results for the $p_T$-integrated flow harmonics of all charged particles, $v_n\equiv \sqrt{\langle v_n^2\rangle}$ (computed within ideal hydrodynamics and averaged over 150 events in each centrality class), and their dependence on $\lambda$. The left panel corresponds to the $0-5\%$ centrality class while the right panel shows the $20-30\%$ centrality results for $\sqrt{s}=2.76$ TeV/n Pb+Pb collisions at the LHC. For the most central collisions, the robustness of the initial eccentricities in Figs.\ \ref{fig:ecc1} and \ref{fig:ecc2} are transferred to the $v_n$'s, which are essentially flat with respect to variations in $\lambda$ from 0.1 to 1 fm. On the other hand, for more peripheral collisions (left panel) we find an enhancement in $v_2$ for larger values of $\lambda$ (this effect is considerably reduced when shear viscosity is taken into account, see \ref{resultshear}). This indicates that, at least in the case of MCKLN initial conditions, $dv_n(\lambda)/d\lambda \gtrsim 0$ and the system's response to the initial spatial eccentricities is somewhat improved when very short wavelength fluctuations are removed and only scales in the mesoscopic regime in Fig.\ \ref{fig:plotscales} are taken into account. It would be interesting to check if this statement holds for other initial state models as well or if it depends on the specific origin of the initial state fluctuations. 

\subsection{Flow coefficients within viscous hydrodynamics and comparison to LHC data}\label{resultshear}

\begin{figure}[ht!]
\centering
\includegraphics[width=0.6\textwidth]{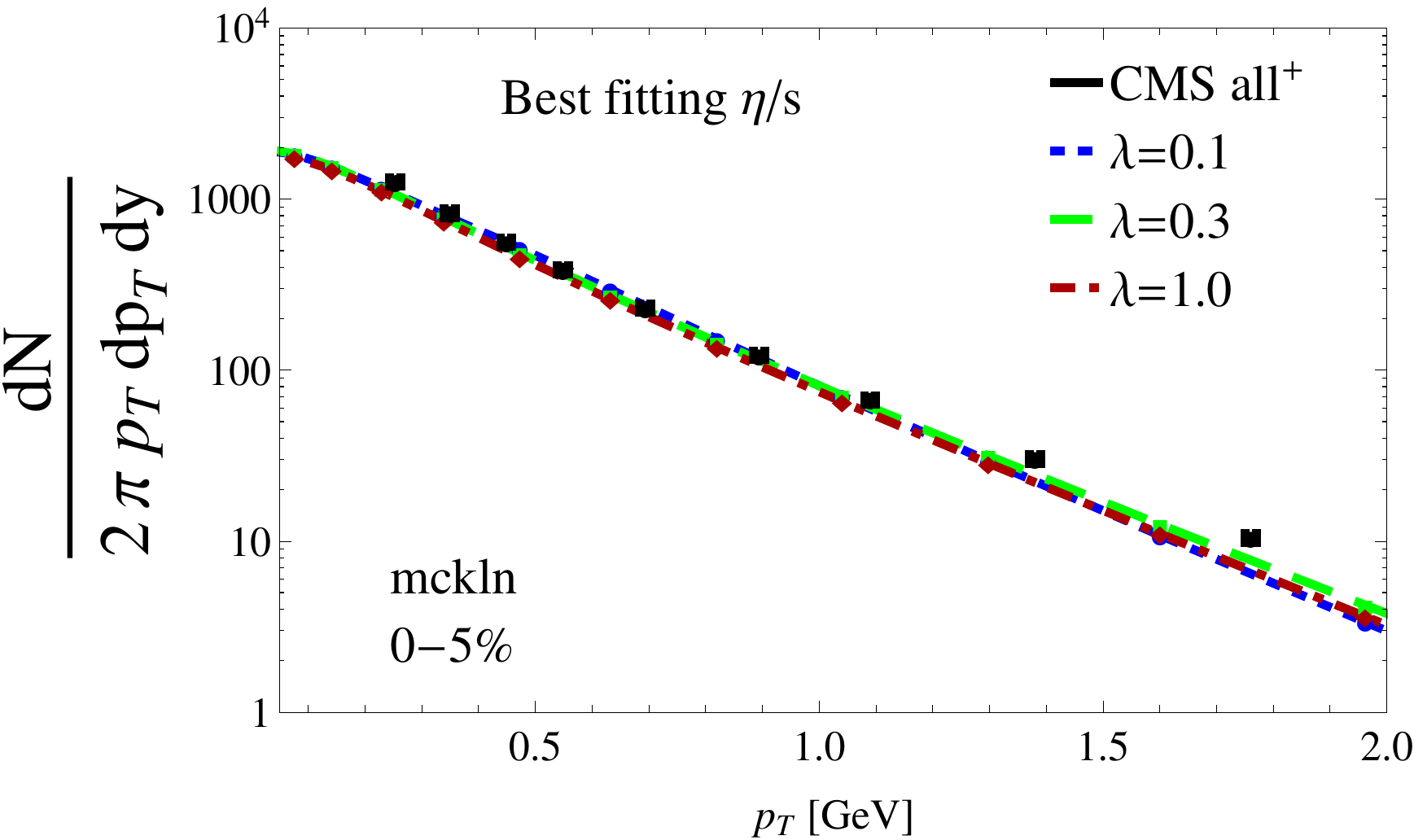}  
\caption{(Color online) $dN/(2\pi p_Tdp_Tdy)$ of all charged particles in the $0-5\%$ centrality class of $\sqrt{s}=2.76$ TeV Pb+Pb collisions at the LHC computed using shear viscous hydrodynamics (averaged over 150 events) with $\lambda=0.3$ fm (and $\eta/s=0.11$) and $\lambda=1$ fm (where $\eta/s=0.1125$). The previous ideal hydrodynamic curve computed with $\lambda=0.1$ fm is also included for comparison. Data points correspond to CMS data \cite{Chatrchyan:2012ta,Chatrchyan:2013kba}.}
\label{fig:specshear}
\end{figure}

In this section we show our viscous hydrodynamic results for $\lambda=0.3$ fm and $\lambda=1$ fm. In order to better describe CMS data, we used $\eta/s=0.11$ for $\lambda=0.3$ fm while for $\lambda=1$ fm we used $\eta/s=0.1125$. In these viscous simulations, the freezeout temperature is $T_{FO}=120$ MeV (unless stated otherwise) and the shear viscous contribution to the distribution function used in the Cooper-Frye freezeout is $\delta f=\pi_{\mu\nu} p^\mu p^\nu/(2 s T^3)$ for all hadronic species. The spectrum of all charged particles computed in the $0-5\%$ centrality class for $\sqrt{s}=2.76$ TeV Pb+Pb collisions at the LHC can be seen in Fig.\ \ref{fig:specshear} where we also included the $\lambda=0.1$ fm ideal hydrodynamic curve from Fig.\ \ref{fig:speclam} for comparison. As before, the data points correspond to CMS data for all charged particles \cite{Chatrchyan:2012ta,Chatrchyan:2013kba}. One can see that the nonzero shear viscosity did not change the qualitative behavior seen already in the ideal case in Fig.\ \ref{fig:speclam}: the low $p_T$ yield is barely affected when one increases $\lambda$ and (small) differences can only be seen above $p_T = 1$ GeV. 

\begin{figure}[ht!]
\centering
\begin{tabular}{cc}
\includegraphics[width=0.4\textwidth]{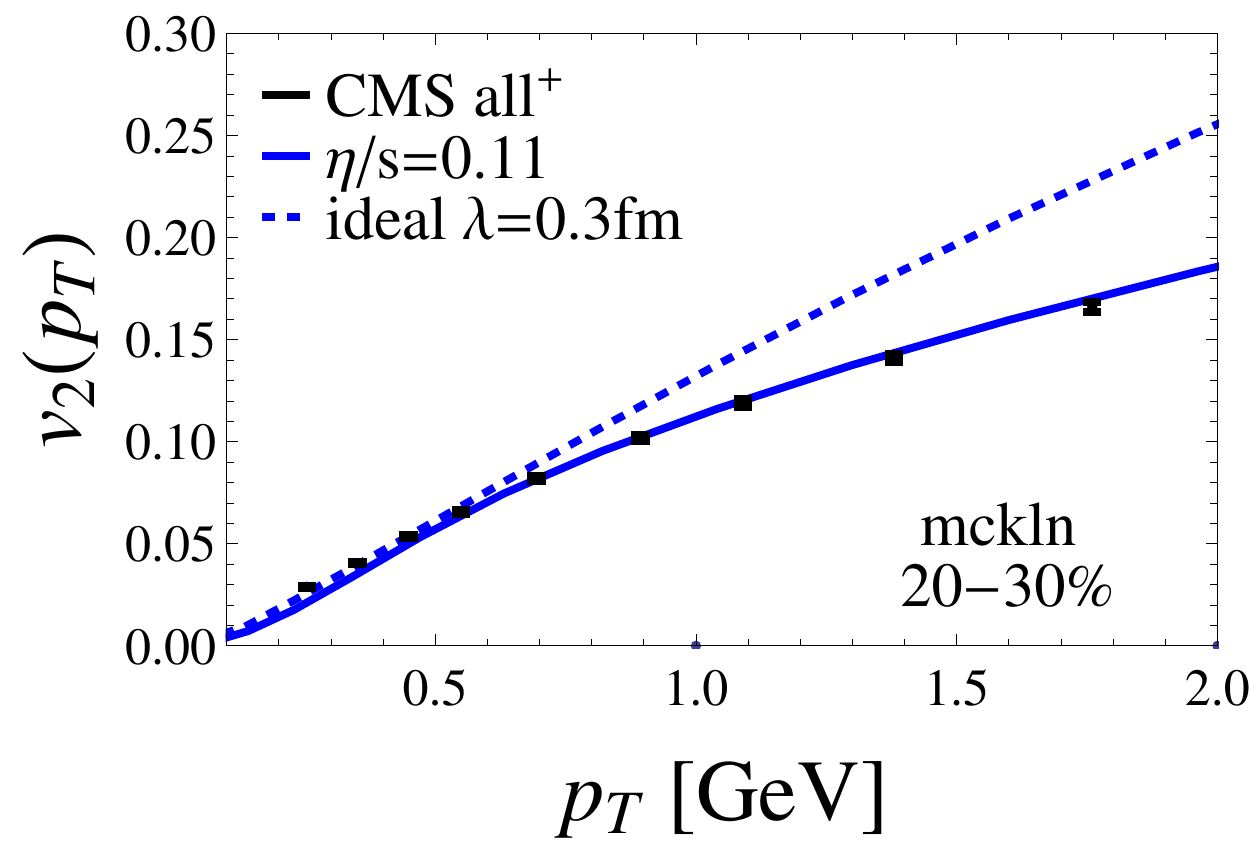} & \includegraphics[width=0.4\textwidth]{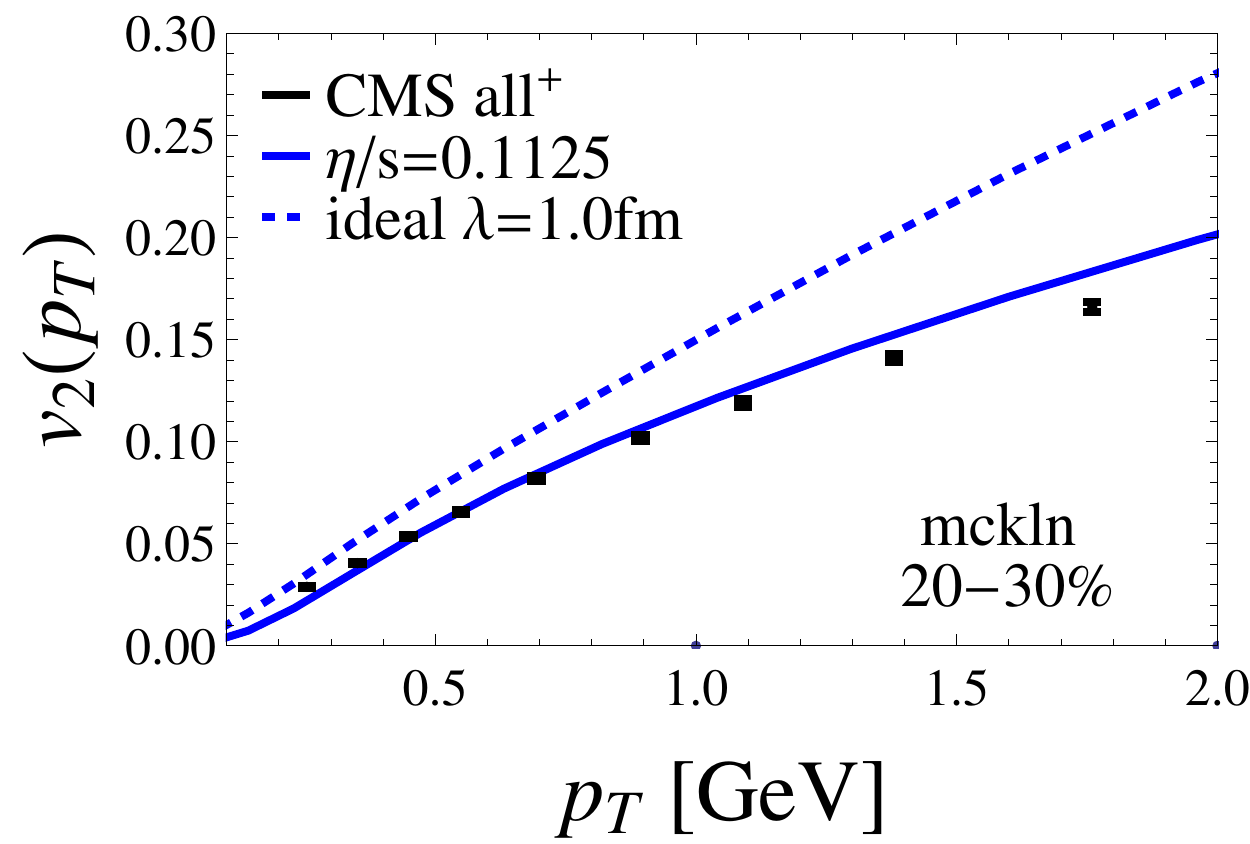} \\
\includegraphics[width=0.4\textwidth]{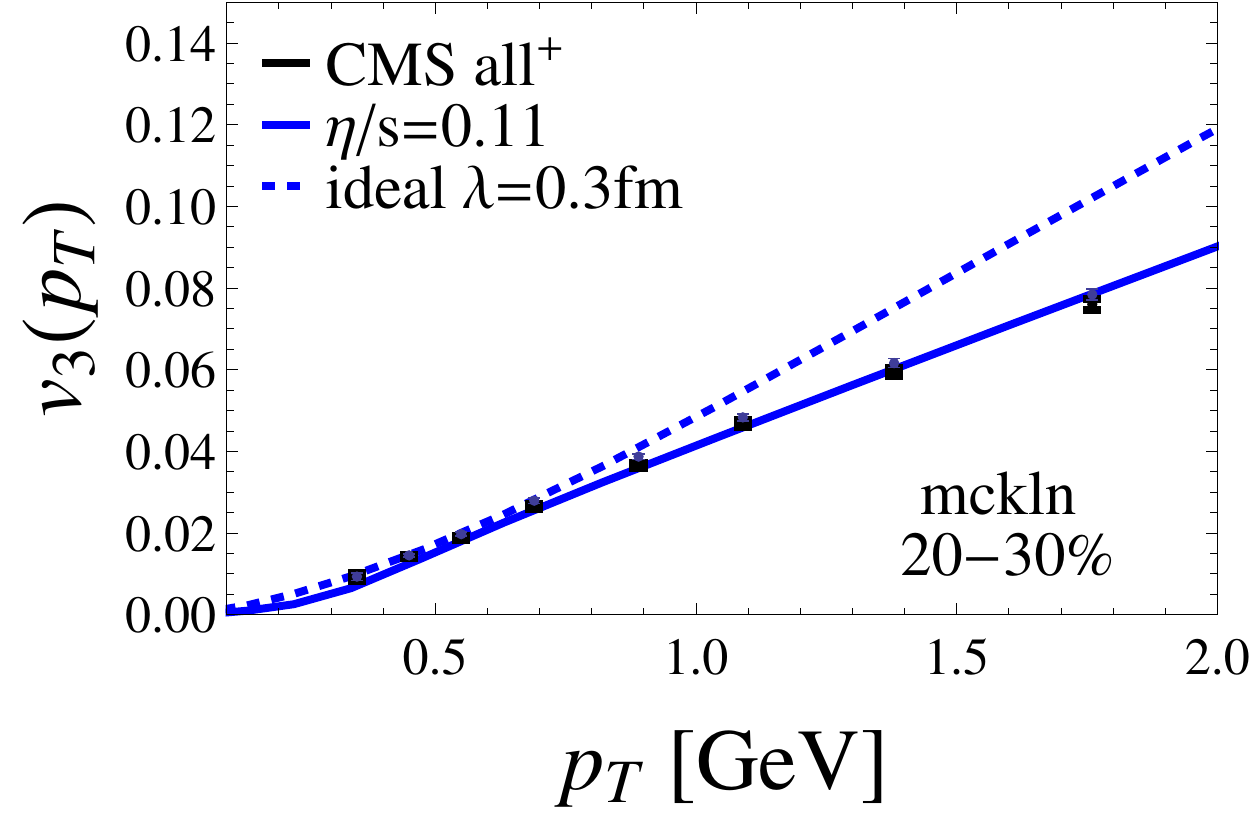} & \includegraphics[width=0.4\textwidth]{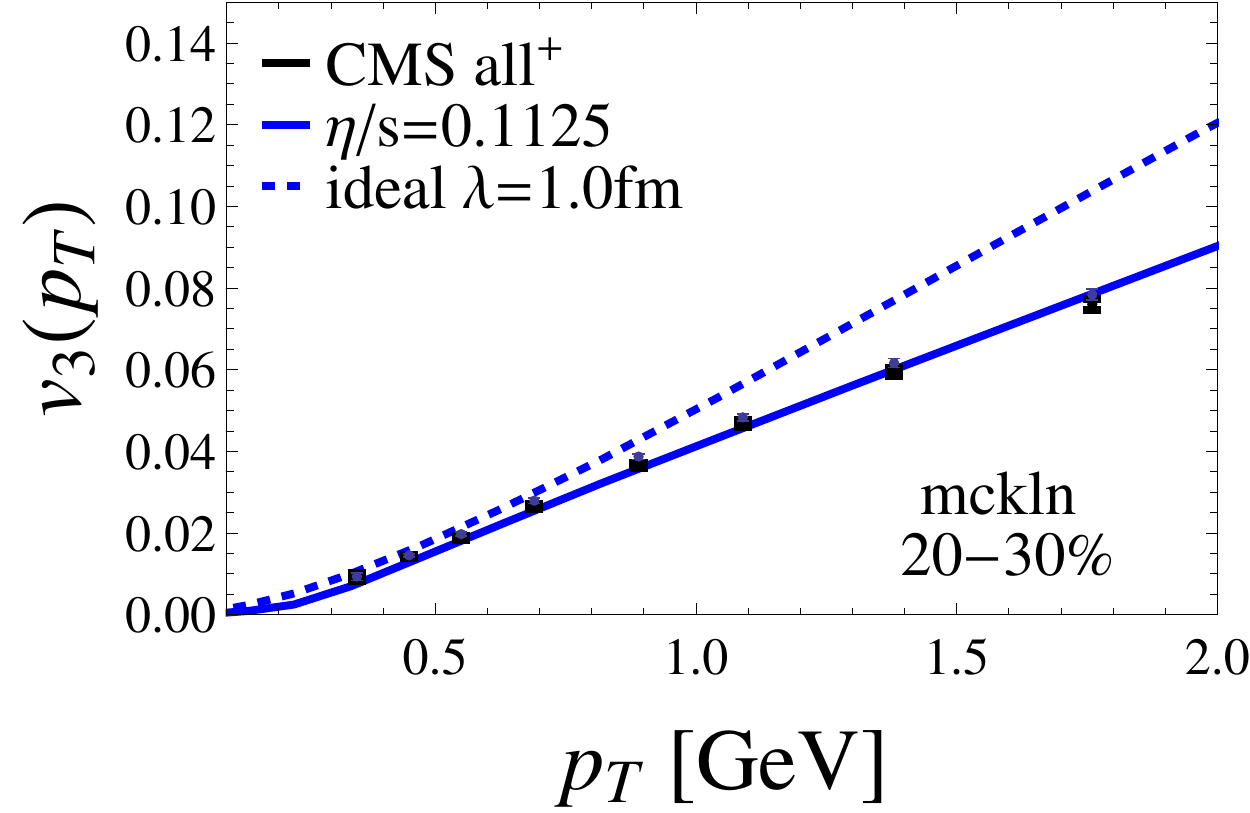} \\
\end{tabular}
\caption{(Color online) Differential flow harmonics $v_2(p_T)$ (top, solid curves) and $v_3(p_T)$ (bottom, solid curves) of all charged particles in the $20-30\%$ centrality class for $\sqrt{s}=2.76$ TeV Pb+Pb collisions at the LHC. The left panels were computed using $\lambda=0.3$ fm while the right panels used $\lambda=1$ fm in the calculations. The data points correspond to CMS data \cite{Chatrchyan:2012ta,Chatrchyan:2013kba} and the solid lines are the shear viscous results for $T_{FO}=120$ MeV. Ideal hydrodynamic results for the flow coefficients from Fig.\ \ref{fig:vnpth} are shown in dashed lines for comparison.}
\label{fig:vnptshear}
\end{figure}

The results for the differential flow harmonics $v_2(p_T)$ and $v_3(p_T)$ for the $20-30\%$ centrality class, computed now within viscous hydrodynamics, can be found in Fig.\ \ref{fig:vnptshear}. One can see that once shear viscosity is included there is almost no difference between the curves computed using $\lambda=0.3$ fm from those where $\lambda=1$ fm. Both values of $\lambda$ give a good description of the CMS data even though the initial energy densities and Knudsen numbers corresponding to these curves (see Figs.\ \ref{fig:edenh} and \ref{fig:knudsen}) reflect very different levels of spatial resolution.

\begin{figure}[ht]
\centering
\includegraphics[width=0.5\textwidth]{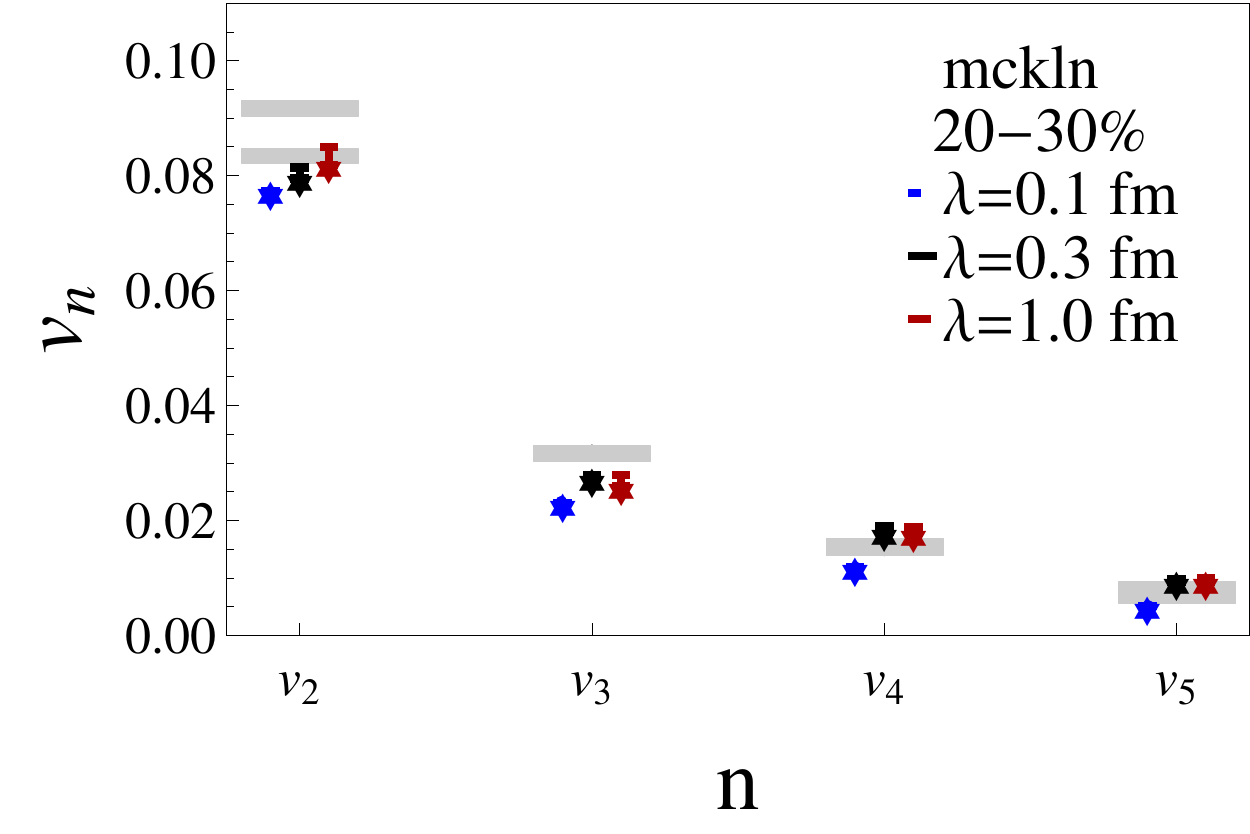} 
\caption{(Color online) Dependence of the integrated flow harmonics $v_n$ of all charged particles (computed within viscous hydrodynamics and averaged over 150 events) on the smoothing scale $\lambda$. The black points were computed using $\lambda=0.3$ fm (with $\eta/s=0.11$) while for the red points $\lambda=1$ fm (and $\eta/s=0.1125$). The small error band attached to these points denote the small dependence of the results when one increases the freezeout temperature from 120 MeV to 130 MeV. The grey bars correspond to CMS data \cite{Chatrchyan:2012ta,Chatrchyan:2013kba} for $\sqrt{s}=2.76$ TeV Pb+Pb collisions at the LHC with the bottom bars denoting the $20-25\%$ centrality while the top bars stand for the $25-30\%$ centrality class (note that for $v_3$, $v_4$, and $v_5$ the bars overlap). We have included the ideal hydrodynamic result for $\lambda=0.1$ fm from Fig.\ \ref{fig:int} for a comparison.}
\label{fig:vnintshear}
\end{figure}

In Fig.\ \ref{fig:vnintshear} we show the corresponding results for the $p_T$ integrated $v_n$'s of all charged particles in the $20-30\%$ centrality class. The grey bars correspond to CMS data \cite{Chatrchyan:2012ta,Chatrchyan:2013kba} with the bottom bars denoting the $20-25\%$ centrality while the top bars stand for the $25-30\%$ centrality class. One can see that the integrated $v_n$'s computed within event-by-event viscous hydrodynamics change very little when $\lambda$ is in the mesoscopic regime. The small error bands attached to the red and black points show how an increase in the freezeout temperature from 120 MeV to 130 MeV affects the $v_n$'s (i.e., they decrease if $T_{FO}$ increases). Also, the viscous hydrodynamic calculations for the flow anisotropies are in the ballpark of the CMS data except for $v_3$ though the ideal hydrodynamic calculation computed using $\lambda=0.1$ fm (the blue points included in the plot for comparison) is below the data for all $n$.   

The results of this section give support to the idea that viscous hydrodynamic calculations of anisotropic flow coefficients are remarkably robust with respect to the spatial resolution scale present in the initial conditions. This shows that the short wavelength fluctuations present in MCKLN initial conditions, which lead to a large Knudsen number in event by event simulations, can be consistently smoothed out without changing significantly the azimuthal anisotropies responsible for generating the flow coefficients $v_n$. It would be interesting to check how the $v_n$ distributions computed within hydrodynamics \cite{Niemi:2012aj,Niemi:2015qia} are affected by variations in $\lambda$. Such a study requires performing hydrodynamic calculations in a number of events an order of magnitude larger than the one presented here and is beyond the scope of this paper.  

\section{Smoothing out proton-nucleus collisions}\label{pAsection}  

We saw in the previous sections that the robustness of the initial eccentricities with respect to variations of the scale of spatial fluctuations in the mesoscopic regime in nucleus-nucleus collisions (across centrality classes) is carried over to the final anisotropic flow coefficients. This is the expected outcome based on the discussion in \cite{Niemi:2012aj}. However, with the advent of the high multiplicity p+Pb  data at the LHC \cite{Chatrchyan:2013nka} and the subsequent observation of significant collective multi-particle behavior in these collisions \cite{Khachatryan:2015waa}, it is interesting to investigate how sensitive the eccentricities and the final flow harmonics in p+Pb collisions are to spatial fluctuations at small scales \cite{Schenke:2014zha}. 

\begin{figure}[ht!]
\centering
\begin{tabular}{c c}
\includegraphics[width=0.45\textwidth]{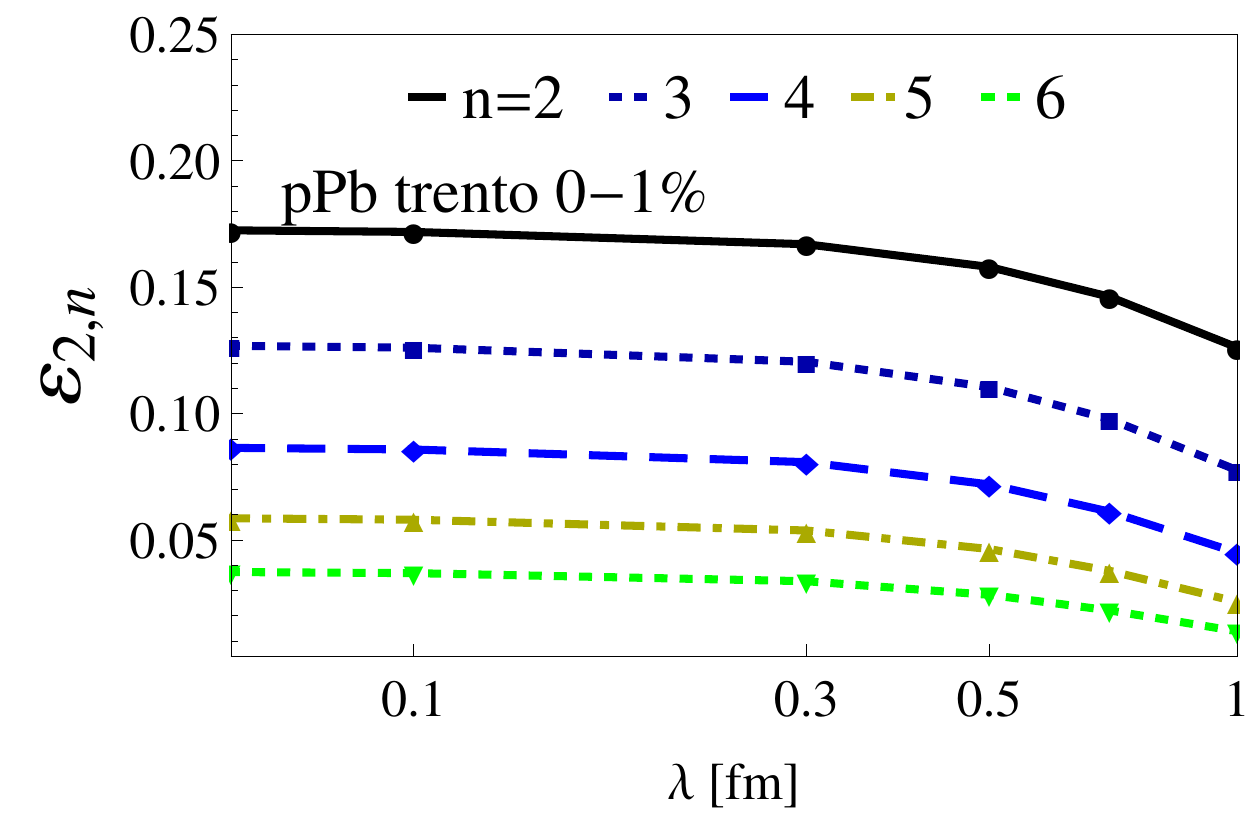} & \includegraphics[width=0.45\textwidth]{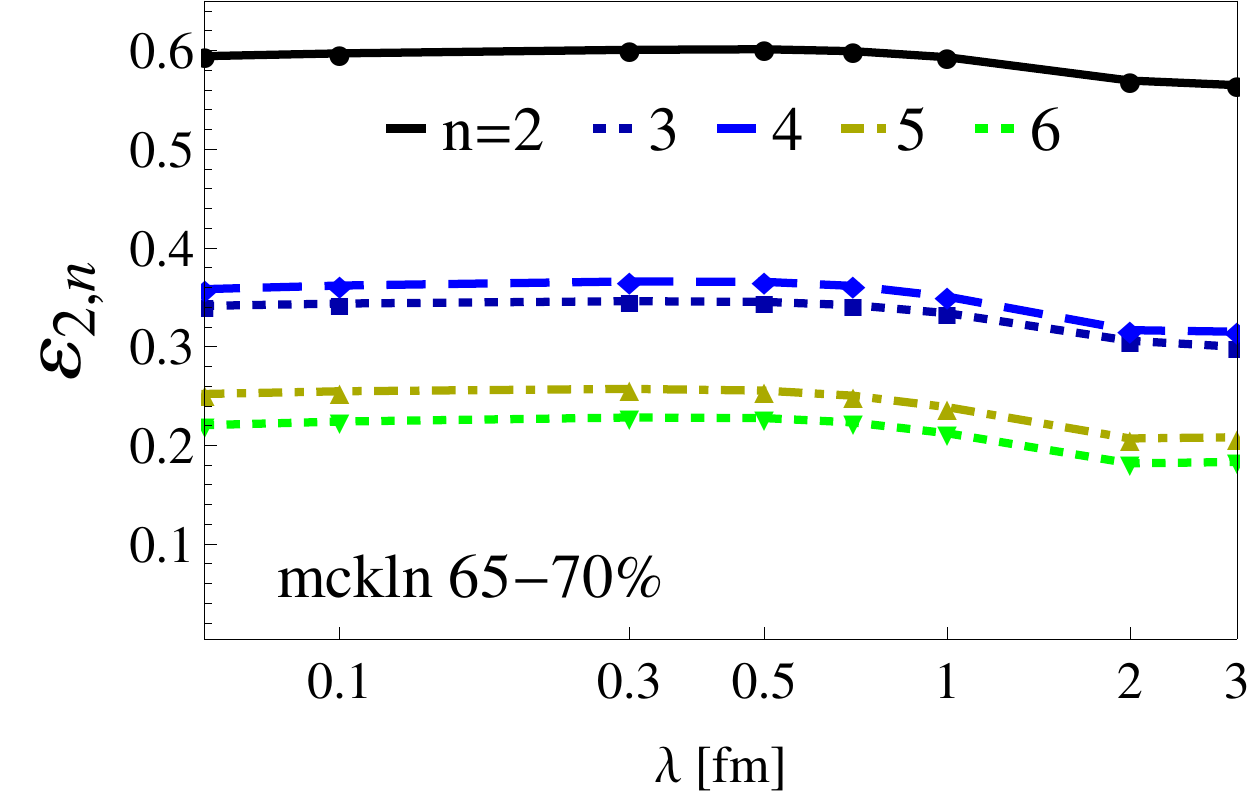} \\
\end{tabular}
\caption{(Color online) Dependence of the initial eccentricities $\varepsilon_{2,n}=\sqrt{\langle \hat\varepsilon_{2,n}^2 \rangle}$ on the smoothing parameter, $\lambda$, for the top $1\%$ high multiplicity $\sqrt{s}=5.02$ TeV p+Pb collisions (left panel) compared to the same quantity in $65-70\%$ centrality class of $\sqrt{s}=2.76$ TeV Pb+Pb events (right panel). These systems have comparable multiplicities \cite{Chatrchyan:2013nka}. The average in each case was performed over 150 events.}
\label{fig:pA}
\end{figure}

In this paper we use the Trento code \cite{Moreland:2014oya} to generate top $1\%$ high multiplicity $\sqrt{s}=5.02$ TeV p+Pb events and check how the eccentricities computed using this model vary with the smoothing parameter. We fix the parameters of the model to be $p=0.3$ and $k=1.3$, which were shown to provide a good description of the number of charged particles per unit rapidity found in different collision systems and centrality classes \cite{Moreland:2014oya}. We show on the left panel in Fig.\ \ref{fig:pA} our results for the $\lambda$ dependence of $\varepsilon_{2,n}$ corresponding to the top $1\%$ high multiplicity $\sqrt{s}=5.02$ TeV p+Pb collisions computed using Trento. On the right panel we show the respective MCKLN results for the $65-70\%$ centrality class of $\sqrt{s}=2.75$ TeV Pb+Pb collisions, which have comparable multiplicities \cite{Chatrchyan:2013nka} to the p+Pb events. One can see that the p+Pb system is much more sensitive to mesoscopic scale fluctuations than peripheral Pb+Pb, even at the same multiplicity. This is to be expected given the smaller size of p+Pb and that in this system the separation of scales in Fig.\ \ref{fig:plotscales} does not apply. Another interesting feature observed in these plots is that the overall magnitude of the eccentricities in p+Pb is a factor of two smaller than those found in peripheral Pb+Pb at the same multiplicity (we note that the p+Pb eccentricities we found using the Trento code are compatible with those found in IP-Glasma \cite{Bzdak:2013zma}). This puts into perspective the experimental finding \cite{Chatrchyan:2013nka} that the two particle cumulant $v_3$ is nearly the same in these systems.

\begin{figure}[ht!]
\centering
\begin{tabular}{c c}
\includegraphics[width=0.45\textwidth]{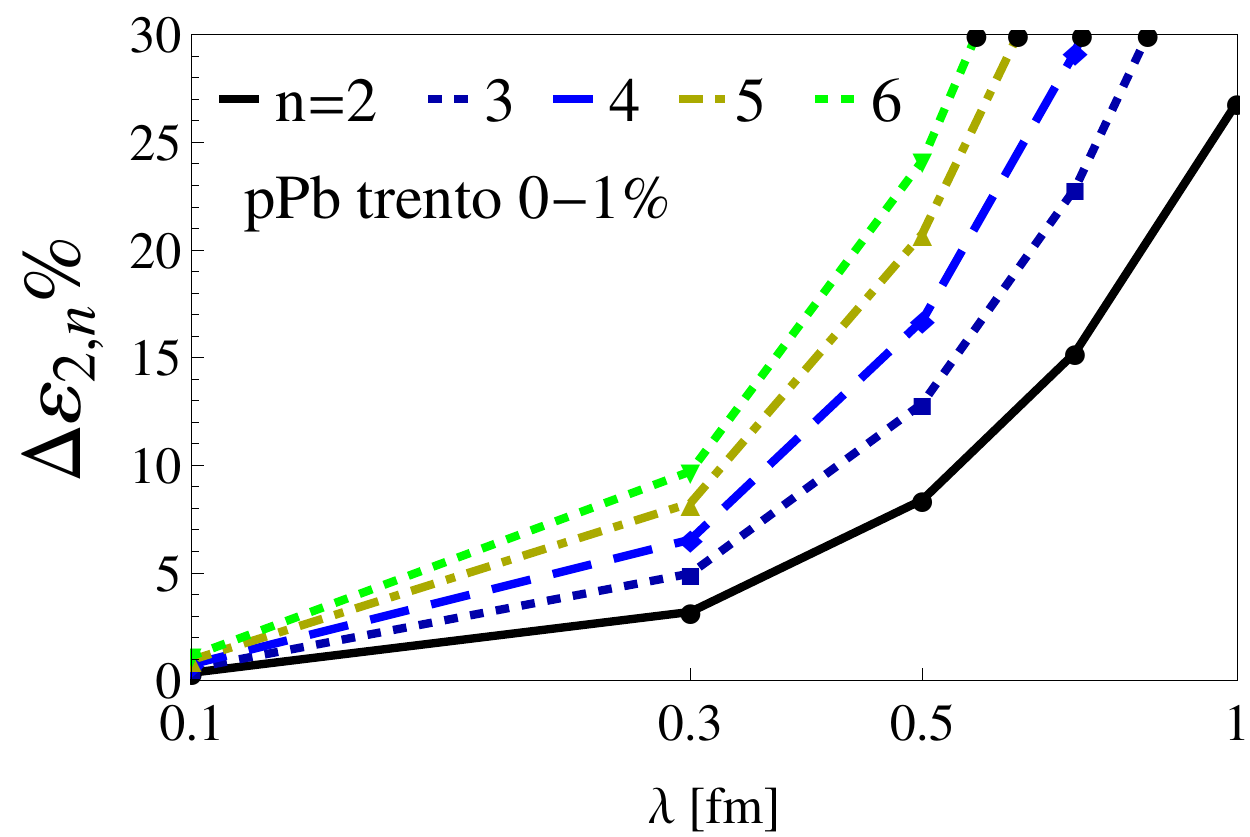} & \includegraphics[width=0.45\textwidth]{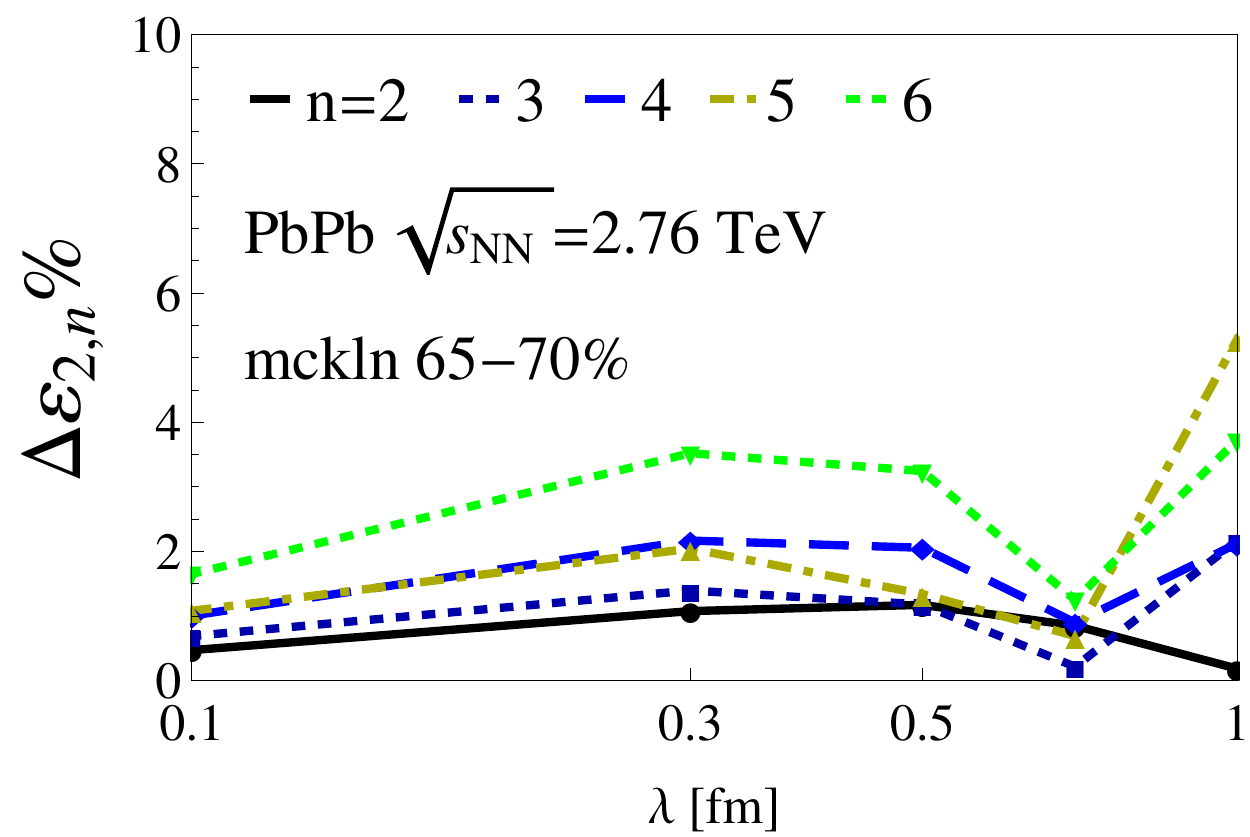} \\
\end{tabular}
\caption{(Color online) Relative variation of initial eccentricities $\Delta \varepsilon_{2,n}=100\left[\varepsilon_{2,n}(\lambda_0)-\varepsilon_{2,n}(\lambda)\right]/\varepsilon_{2,n}(\lambda_0)$ on the smoothing parameter, $\lambda$, for the top $1\%$ high multiplicity $\sqrt{s}=5.02$ TeV p+Pb collisions (left panel) compared to the same quantity in $65-70\%$ centrality class of $\sqrt{s}=2.76$ TeV Pb+Pb events (right panel). The average in each case was performed over 150 events.}
\label{fig:pA2}
\end{figure}

The fact that p+Pb is much more sensitive to the smoothing parameter than peripheral Pb+Pb can be more clearly seen in Fig.\ \ref{fig:pA2} where we show the relative variation of the eccentricities with $\lambda$. While in peripheral Pb+Pb the variation of the eccentricities remain below $10\%$ for  $\lambda=0.1-1$ fm, the same cannot be said about p+Pb where $10\%$ variations are already found in the interval $\lambda=0.1-0.3$ fm (note, however, that the variations in the lowest order eccentricities $\varepsilon_{2,2}$ and $\varepsilon_{2,3}$ remain $\sim 10\%$ up to $\lambda=0.5$ fm). We note here that if one uses the Trento setup specific to LHC found in \cite{Moreland:2014oya} of  $p=0$ and $k=1.4$ that the initial eccentricities are even smaller and even more sensitive to the smoothing scale and even see large deviations for  $\varepsilon_{2,2}$ and $\varepsilon_{2,3}$ above $\lambda=0.3$. This shows that p+Pb collisions should be very sensitive to the underlying physics below the confinement scale, which may justify the search for alternative non-hydrodynamical explanations for the p+Pb flow harmonics data \cite{Dusling:2012wy,Noronha:2014vva,Gyulassy:2014cfa,Dumitru:2014dra,Dumitru:2014yza,Lappi:2015vha,Schenke:2015aqa}.      

\section{Conclusions and Outlook}\label{conclusions}

We argued in this paper that QCD phenomena in ultrarelativistic collisions of heavy ions may be separated into three distinct regimes, illustrated in Fig.\ \ref{fig:plotscales}. The microscopic sub-nucleon scale regime involves phenomena that are sensitive to length scales of the order of $1/Q_s$ while one may take $1/\Lambda_{QCD}$ as a mesoscopic scale above which one encounters a macroscopic nuclear scale regime given by length scales of the order of the radius of a large nucleus. Though there is currently no single effective theory that is able to describe all the different phenomena that occur at these different length scales, the current success of viscous hydrodynamics in understanding several features of heavy ion data implies that it is should be a good starting point to describe the bulk dynamics in the mesoscopic regime since that is actually probed in event by event calculations. In this case, it is important to remark that current viscous hydrodynamic implementations based on Israel-Stewart hydrodynamics do not have the correct degrees of freedom to describe scales shorter than the respective relaxation time (in our case, $\tau_\pi$), which provides a lower bound on the length scales at which these hydrodynamic models are sensible. The presence of these different scales in the description of ultrarelativistic heavy ion collisions calls for a systematic study of the effects of initial state fluctuations, separated according to their wavelength \cite{Floerchinger:2013rya}, on hydrodynamically generated observables such as the anisotropic flow coefficients.

In this paper we used a Lagrangian based procedure to smooth out the initial conditions employed in hydrodynamics simulations of ultrarelativistic heavy ion collisions that allowed us to investigate, in a systematic manner, how the final azimuthal anisotropies $v_n$'s (computed within event by event hydrodynamics) depend on the wavelength of the initial energy density fluctuations. The smoothing procedure is characterized by a smearing function with a finite support given by the smoothing scale $\lambda$ that defines the maximal size of the initial energy density gradients. The method can be applied in a variety of initial condition models ranging from MCKLN/MCGlauber to IP-Glasma. Here we used $\sqrt{s}=2.76$ TeV Pb+Pb MCKLN initial conditions to show that the initial spatial eccentricities that govern the final state flow harmonics are very robust with respect to variations in the underlying scale of initial energy density fluctuations (or smoothing parameter) $\lambda$, as long as these fluctuations are in the so-called mesoscopic regime (see Fig.\ \ref{fig:plotscales}) where $0.1<\lambda <1$ fm.   

Given that the local Knudsen number is $K_n \sim 1/\lambda$, the robustness of the initial eccentricities with respect to changes in the fluctuation scale in the mesoscopic regime was carried over to the $v_n$'s computed within viscous hydrodynamics. This implies that events with large local $K_n$ and events where $K_n$ is near the hydrodynamic regime lead to nearly indistinguishable flow harmonics. In fact, at least in the case of Pb+Pb MCKLN initial conditions (and most certainly also for MCGlauber), the anisotropic flow coefficients computed within event by event viscous hydrodynamics are only sensitive to long wavelength scales of order $ 1/\Lambda_{QCD}\sim 1$ fm and are, thus, very robust with respect to variations in the initial local Knudsen number.

Since the structures at very small scales present in MCKLN initial conditions are still invariably rooted in the MCGlauber-like fluctuations of the positions of the (independent) nucleons in the nucleus, one may say that $ 1/\Lambda_{QCD}$ should indeed be the upper limit of the mesoscopic regime in our case. Therefore, it is important to investigate if the conclusions found here regarding the robustness of the flow harmonics in viscous hydrodynamics in nucleus-nucleus collisions still hold when the short wavelength fluctuations in the initial energy density stem from nontrivial color fields defined in the microscopic sub-nucleon scale regime $\sim 1/Q_s$ shown in Fig.\ \ref{fig:plotscales}. One should keep in mind, however, that taking into account fluctuations of very short wavelength scales considerably worsens the local Knudsen number (see Fig.\ \ref{fig:knudsenbad}). Moreover, since the relaxation time provides a lower limit on the scales that can be described within Israel-Stewart hydrodynamics, in order to assess very short wavelength fluctuations characterized by $\lambda < 0.1$ fm, $\tau_\pi$ would have to be significantly small.

Furthermore, effective theory arguments may be used to argue that truly microscopic sub-nucleon $\sim 1/Q_s$ physics should not influence the bulk dynamics of matter at much larger length scales $\sim 1/\Lambda_{QCD}$, which seem to be the relevant ones for anisotropic flow coefficients in nucleus-nucleus collisions where the separation of scales shown in Fig.\ \ref{fig:plotscales} approximately holds. However, it is known that in many-body systems sometimes the existence of a large separation of scales is not sufficient to rule out microscopic effects\footnote{The stability of matter (see \cite{lieb}) relies on the Pauli principle, which is not only essential in the description of matter in subatomic scales but is also ultimately fundamental to ensure the stability of macroscopic bodies.} and, thus, it would definitely be interesting to check how the anisotropic flow generated with other type of initial conditions is affected by the smoothing procedure defined in \ref{smoothingsection1}. Also, it would be interesting to understand how fluctuations in the sub-nucleon regime may affect the approximate scaling properties found in heavy ion observables discussed in \cite{Torrieri:2013aqa,Basar:2013hea} and also how they can influence a principal component analysis \cite{Bhalerao:2014mua,Mazeliauskas:2015vea}. Furthermore, we intend to investigate the case of nucleus-nucleus collisions with reduced center of mass collision energies $\sqrt{s}$, such as in the low energy beam energy scan program at RHIC.

We also investigated how the eccentricities of the top $1\%$ high multiplicity p+Pb collisions are affected by our smoothing procedure. In this case, the scenario depicted in Fig.\ \ref{fig:plotscales} does not hold since the separation between the macroscopic and the mesoscopic nuclear regimes becomes artificial and much less motivated than in the case of nucleus-nucleus collisions. We found that the eccentricities in p+Pb are very sensitive to sub-nucleon scale fluctuations, which should be contrasted with the robustness found in peripheral Pb+Pb collisions with the same multiplicity.

Also, regarding the hydrodynamic evolution, it is important to take into account the temperature dependence of transport coefficients such as $\eta/s$ (and $\zeta/s$) since that will affect our results for the Knudsen number \cite{Niemi:2014wta}. The inclusion of other second order hydrodynamic coefficients beyond $\tau_\pi$ in the hydrodynamic evolution may also play a role in determining $K_n$, especially in small systems. Moreover, it would be interesting to extend the study performed here to the case of 3+1 viscous hydrodynamic simulations including the possibility of different smoothing scales in the transverse and longitudinal directions.

Since the underlying bulk dynamics of the expanding medium is relevant for energy loss calculations \cite{Burke:2013yra}, it would be interesting to check to what extent jet quenching-related observables are sensitive to short wavelength fluctuations. An initial study performed in \cite{Betz:2011tu} showed little sensitivity to initial state fluctuations, though it should be kept in mind that a full event-by-event hydrodynamic evolution of the initial spatial inhomogeneities was not included in their analysis. Additionally, one may also study how recent calculations of the medium response to the energy and momentum lost by jets \cite{Tachibana:2014lja,Andrade:2014swa,Schulc:2014jma} are affected by different levels of spatial resolution parametrized in terms of the smoothing scale $\lambda$.

\acknowledgments

We thank T.~Kodama for general discussions about the coarse-graining process in heavy ion collisions, A.~Dumitru and G.~S.~Denicol for enlightening discussions about the effects of sub-nucleon initial state fluctuations in hydrodynamics, G.~Torrieri for suggesting to check the role of fluctuations in different collision systems and energies, M.~Luzum for discussions on the hydrodynamic approach to proton-nucleus collisions, and S.~Moreland and J.~Bernhard for assistance with using the Trento initial condition code. JNH and MG acknowledge support from the US-DOE Nuclear Science Grant No. DE-FG02-93ER40764. JN thanks the Physics Department at Columbia University for its hospitality and Funda\c c\~ao de Amparo \`a Pesquisa do Estado de S\~ao Paulo (FAPESP) and Conselho Nacional de Desenvolvimento Cient\'ifico e Tecnol\'ogico (CNPq) for support.

\end{document}